\RequirePackage{fix-cm}
\documentclass[twocolumn,epjc3]{svjour3}  
\smartqed  
\RequirePackage{graphicx}

\usepackage{amsmath}
\usepackage{graphicx}
\usepackage{dcolumn}
\usepackage{bm}
\usepackage{soul}

\usepackage{epsfig}
\usepackage{dcolumn}
\usepackage{float}
\usepackage[]{hyperref}
  \hypersetup{
  unicode=false,          
  pdftoolbar=true,        
  pdfmenubar=true,        
  pdffitwindow=true,     
  pdfstartview={FitH},    
  pdfsubject={On the Optimization of Protvino to ORCA (P2O) Experiment the DUNE early 
physics output with current experiments},   
  pdfnewwindow=true,      
  pdfcreator={RevTeX},
  colorlinks=true,       
  linkcolor=red,          
  citecolor=blue,        
  urlcolor=blue,           
  }
\usepackage{hypcap}
\usepackage[utf8]{inputenc}

\def\be{\begin{equation}}
\def\ee{\end{equation}}
\def\bea{\begin{eqnarray}}
\def\eea{\end{eqnarray}}
\def\gsim{\ \rlap{\raise 2pt\hbox{$>$}}{\lower 2pt \hbox{$\sim$}}\ }
\def\lsim{\ \rlap{\raise 2pt\hbox{$<$}}{\lower 2pt \hbox{$\sim$}}\ }
\def\dslash{\kern-4pt \not{\hbox{\kern-2pt $\partial$}}}
\def\pslash{\not{\hbox{\kern-2pt p}}}

\newcommand{\dcp}{\delta_{CP}}

\newcommand{\cnv}{\v{C}erenkov}

\journalname{Eur. Phys. J. C}
\begin{document}

\title{Sensitivity study of Protvino to ORCA (P2O) experiment: Effect of antineutrino run, background and systematics 
}


\author{Sandhya Choubey\thanksref{e1}
        \and
        Monojit Ghosh\thanksref{e2} 
        \and
        Dipyaman Pramanik\thanksref{e3} 
}

\thankstext{e1}{e-mail: sandhya@hri.res.in}
\thankstext{e2}{e-mail: ghoshmonojit@hri.res.in}
\thankstext{e3}{e-mail: dipyamanpramanik@hri.res.in}


\institute{Harish-Chandra Research Institute, HBNI, Chhatnag Road, Jhunsi, Allahabad 211 019, India}

\date{Received: date / Accepted: date}

\maketitle

\begin{abstract}
There is a proposal to send a neutrino beam from the Protvino accelerator complex located in Russia to the detector facility called `Oscillation Research with Cosmics in the Abyss' (ORCA)
in the Mediterranean sea 
to study neutrino oscillation. This is called the P2O experiment which will have a baseline of 2588 km. In this paper we carry out an sensitivity study to extract the best possible physics sensitivity of the P2O experiment. In particular, we study the effect of antineutrino runs, the role of background as well as the impact of controlling the systematic uncertainties vis-a-vis the statistics. 

\end{abstract}

\section{Introduction}
\label{intro}

Ambitious and expensive next-generation long-baseline experiments are being proposed to measure the remaining crucial ingredients of the neutrino mass matrix - (i) the sign of $\Delta m_{31}^2$ {\it aka} the neutrino mass hierarchy, (ii) the octant of $\theta_{23}$ and (iii) CP violation. The Deep Underground Neutrino Experiment (DUNE) \cite{Acciarri:2015uup} has been proposed to be built in USA, Tokai to Hyper-Kamiokande (T2HK) \cite{Abe:2016ero} proposal has been made for a long-baseline experiment in Japan and ESS$\nu$SB \cite{Baussan:2013zcy} has been put forth as a prospective long-baseline endeavour in Sweden. The detector technology for DUNE will be liquid argon time projection chamber, while for T2HK and ESS$\nu$SB water \cnv\ detector will be used. The CP sensitivity of all these three proposals is very high. DUNE is designed to measure CP violation at around $5 \sigma$ C.L. with its proposed 3+3 years run for maximal CP violation\footnote{By $x+y$ we imply running the experiment for $x$ years in the neutrino mode and $y$ years in antineutrino mode.}. T2HK should be able to discover CP violation at $7\sigma$ C.L. in 5+5 years (assuming that the mass hierarchy is known), while ESS$\nu$SB's projected sensitivity to CP violation is $8\sigma$ from a 5+5 years run. The octant sensitivity at these experiments is also seen to be very good with DUNE, T2HK and ESS$\nu$SB promising a $3\sigma$ discovery for ($\theta_{23}<43.5^\circ$ and $\theta_{23}>47.5^\circ$), ($\theta_{23}<43^\circ$ and $\theta_{23}>48^\circ$) and ($\theta_{23}<41^\circ$ and $\theta_{23}>50^\circ$), respectively. On the other hand, the hierarchy sensitivity is good only in DUNE where we can expect a $15\sigma$($7.5\sigma$) discovery for $\delta_{CP}=-90^\circ(+90^\circ)$\footnote{Throughout the paper we use the acronyms NH for normal hierarchy and IH for inverted hierarchy.}  after a total run-time of 10 years. The ESS$\nu$SB  set-up is not expected to have any hierarchy sensitivity at all, while for T2HK little hierarchy sensitivity is expected if both the water tanks are placed in Japan at a distance of 295 km. In order to alleviate this problem to some extent, there is a proposal  to put the second tank of the Hyper-Kamiokande detector in Korea, at a distance of about 1100 km from Tokai \cite{Abe:2016ero}, bringing in matter effects and hence some hierarchy sensitivity. This proposal is called Tokai to Hyper-Kamiokande Korea (T2HKK) \cite{Abe:2016ero}. The hierarchy sensitivity for T2HKK proposal though is expected to be not significantly above $5\sigma$ C.L. even after 5+5 years of run time. Note that all the projected sensitivity reaches mentioned above are assuming NH to be true. For IH the sensitivities of all experiments are expected to be lower, especially for mass hierarchy. For studies regarding measuring the unknowns in the standard three flavour framework in these future long-baseline experiments see Ref. \cite{Agarwalla:2017nld,Ghosh:2017ged,Fukasawa:2016yue,Nath:2015kjg,Ghosh:2014rna,Chakraborty:2017ccm,Raut:2017dbh,Barger:2014dfa,Barger:2013rha,Agarwalla:2014tpa,Chatterjee:2017irl,Chatterjee:2017xkb,Blennow:2014sja,Ballett:2016daj,Evslin:2015pya}.

The main reason for lower hierarchy sensitivity in the above mentioned next-generation long-baseline experiments is their shorter baseline. To keep the experiment at the oscillation maximum the corresponding fluxes are made to peak at lower energies such that $L/E$ remains constant. The matter potential of the earth is proportional to the energy of the neutrino. Therefore  the matter potential is smaller for the shorter baseline experiments and thus the hierarchy sensitivity is smaller in these experiments. 
To overcome this, there is another European experimental proposal with a relatively longer baseline. This proposal is known as the Protvino to ORCA (Oscillation Research with Cosmics in the Abyss) or the P2O experiment \cite{p2o_talk_zaborov,Brunner:2013lua,Zaborov:2018whl,Adrian-Martinez:2016fdl,Akindinov:2019flp}. This proposal comprises of shooting a neutrino beam from the Protvino accelerator site in Russia to the ORCA detector in the Mediterranean sea at a distance of 2588 km from the accelerator. The Protvino accelerator site currently houses the U-70 accelerator. The proposal is to  upgrade the accelerator with an increased beam power of 450 KW to deliver $4\times 10^{20}$ protons on target per year. These protons are collided with a target to produce pions. These pions then go through a focusing system and decay pipe to produce a neutrino beam which is peaked at about 5 GeV, in order to coincide with the oscillation maximum for this baseline \cite{p2o_beam}. This larger beam energy brings in significant matter effects in the oscillation probability giving P2O a clear edge over the other next-generation experiments in measuring the neutrino mass hierarchy. We will see that even with just one year of running of this experiment, one would measure the neutrino mass hierarchy at more than $3.5\sigma$ in the worst possible scenario and greater than $5\sigma$ in the best possible case. This extremely high sensitivity of P2O to neutrino mass hierarchy comes also from the megaton-scale mass of the ORCA detector as well as the fact that the 2588 km baseline is very close to being bi-magic \cite{Dighe:2010js,Raut:2009jj}\footnote{ At bi-magic basline ($L = 2540$ km), there is no $\dcp$ dependence in NH (IH) for 1.9 (3.3) GeV. This ensures hierarchy measurement without any $\dcp$ degeneracy with a high confidence level.}. 

In this paper we make a detailed study on the sensitivity of the experimental set-up for measuring the three unknowns in neutrino physics mentioned above. We will look into how the detector systematics and detector backgrounds affect the CP violation and octant of $\theta_{23}$ sensitivities of this experiment vis-a-vis the number of years of running of the experiment. We find that the hierarchy sensitivity of P2O, especially for NH true, is so high that any deterioration coming from uncontrolled backgrounds and/or systematic uncertainties makes no practical difference and one is ensured a sure shot discovery of the neutrino mass hierarchy within a very short time. The impact of systematics and backgrounds show up when the hierarchy is inverted, nonetheless a clear discovery of mass hierarchy is guaranteed, albeit with slightly higher run time than needed for the NH true case. We will show that the time needed for hierarchy discovery at P2O even with IH true will be significantly shorter than that needed at other competing experiments.  On the other hand, the CP violation and octant sensitivity of P2O is seen to be weaker than the corresponding expected sensitivity at competing long-baseline experiments. We will show that these sensitivities can be improved mildly to severely with changes in systematic uncertainties and background. Also, for all cases, we study the impact of adding the antineutrino run to the projected sensitivity at P2O. Note that the P2O collaboration in their first results \cite{p2o_talk_zaborov} has shown the sensitivity to hierarchy and CP violation using 3 years run in the neutrino channel only. We find that a small antineutrino run fraction helps in getting rid of parameter degeneracies and helps in improving the expected sensitivity of CP violation and octant of $\theta_{23}$. We will quantify each of these aspects in what follows.

The rest of the paper is organised in the following way. In section \ref{simulation} we spell out our experimental and simulation details. In section \ref{results} we present the main results of our paper assuming NH to be true. In section \ref{inverted} we present our main findings on what happens when the true hierarchy is inverted instead of normal. Finally in section \ref{comparison} we compare the sensitivity of the P2O experiment with the other future long-baseline experiments. We conclude in section \ref{conclusions}.

\begin{figure*}[htb]
\begin{center}
\includegraphics[width=0.56\textwidth]{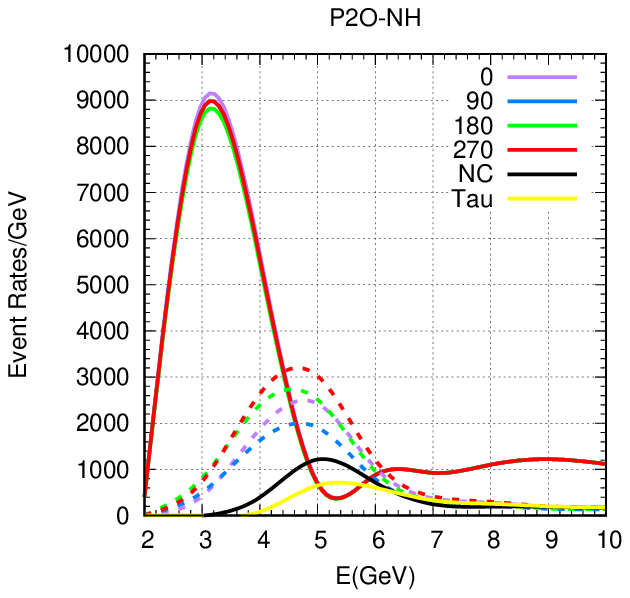} 
\hspace{-1.0 in}
\includegraphics[width=0.56 \textwidth]{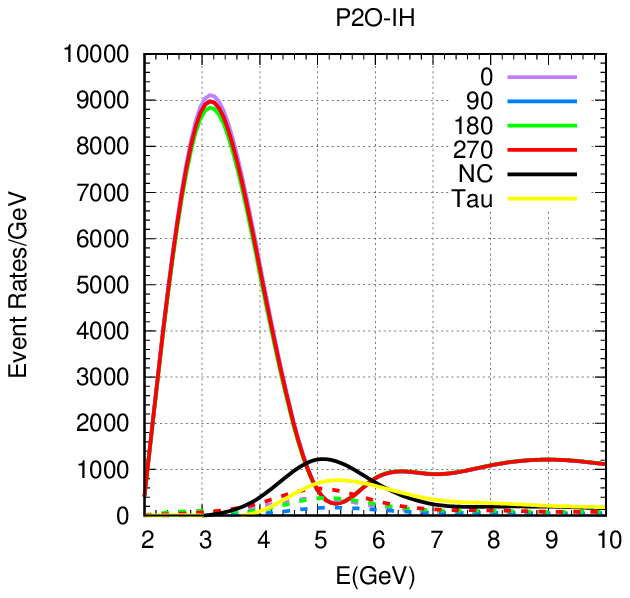}
\caption{Events spectrum vs energy. Numbers are generated at $\theta_{12} = 33.4^\circ$, $\theta_{13} = 8.42^\circ$, $\Delta m^2_{21} = 7.53 \times 10^{-5}$ eV$^2$ and $\Delta m^2_{31} = \pm 2.44 \times 10^{-3}$ eV$^2$. The solid (dashed) purple/blue/green/red curves correspond to muon (electron) events for $\dcp=0^\circ/90^\circ/180^\circ/270^\circ$. These events corresponds to three years running of P2O in the neutrino mode corresponding to $12 \times 10^{20}$ proton on target.}
\label{events}
\end{center}
\end{figure*}

\begin{figure*}[htb]
\includegraphics[width=0.45\textwidth]{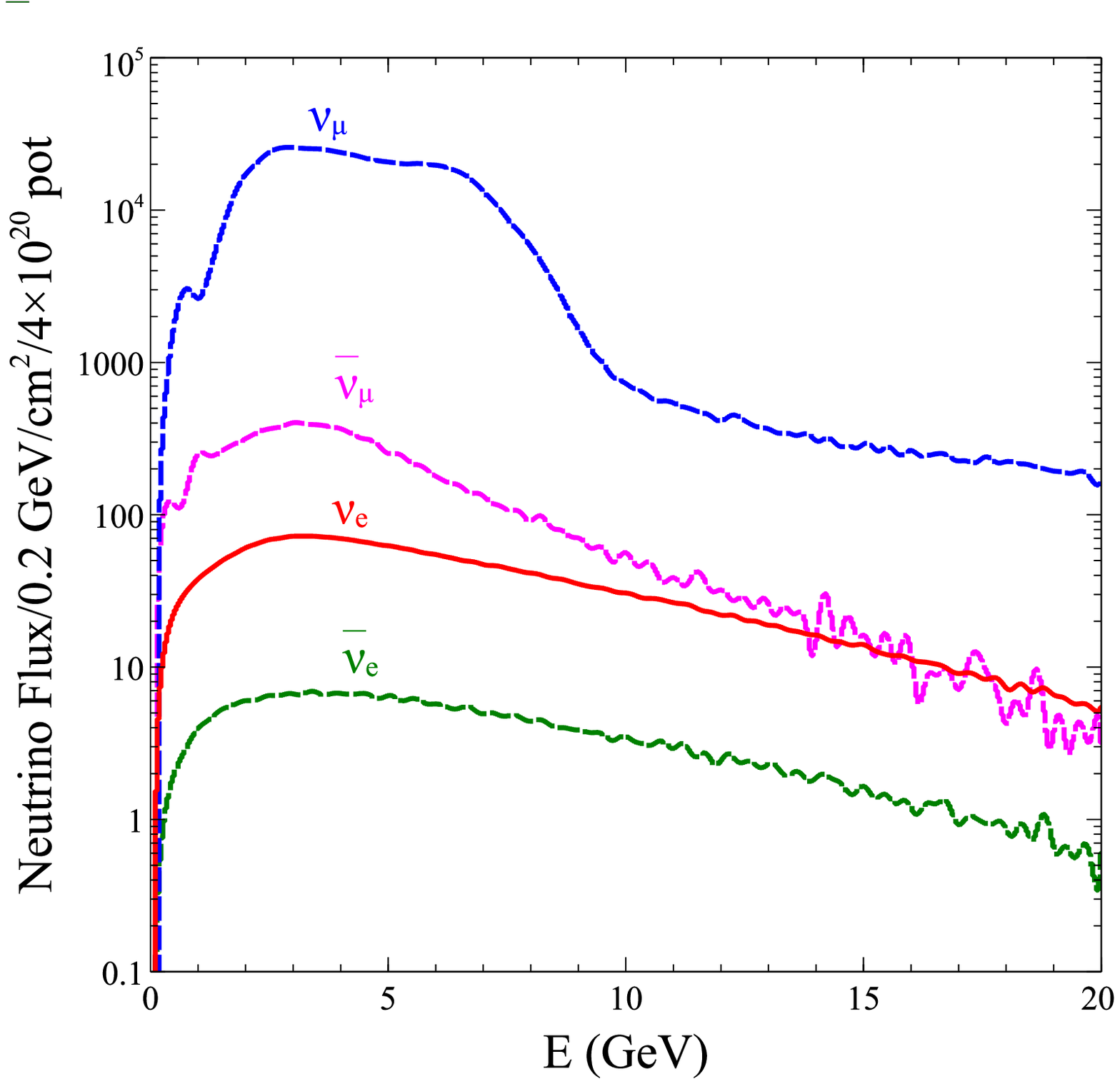} 
\hspace{0.5 in}
\includegraphics[width=0.45\textwidth]{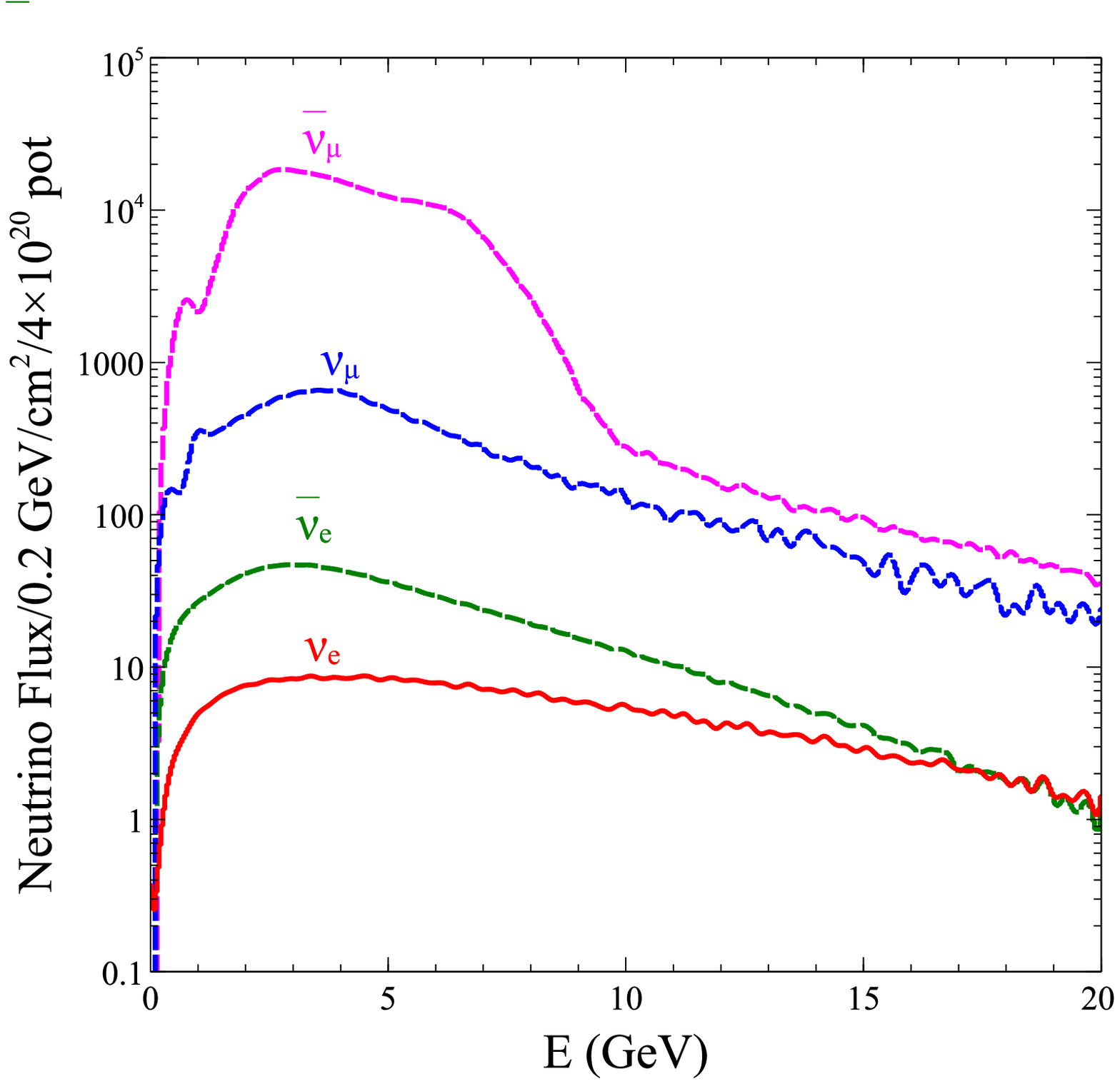}
\caption{Fluxes as a function of energy. The left panel is for neutrino flux and the right panel is for antineutrino flux.}
\label{flux}
\end{figure*}

\section{Experimental and Simulation Details} 
\label{simulation}

As mentioned above, the U-70 accelerator located at the Protvino accelerator complex 100 km south of Moscow,  is proposed to be upgraded to have 450 KW beam power to deliver $4 \times 10^{20}$ protons on target (pot) per year. The neutrinos produced in this accelerator will be detected at ORCA at a distance of 2588 km. At this baseline the first oscillation maximum occurs at 5 GeV in vacuum. To simulate the P2O experiment we follow the experimental details given in Ref \cite{p2o_talk_zaborov}\footnote{Note that while our paper was under review, a detailed analysis by the P2O collaboration appeared on arXiv \cite{Akindinov:2019flp} which mentioned that the results found are comparable to ours.}. Our first aim is to obtain the event rates $N$. We take 1 GeV bins in the energy range 1.5 GeV to 10 GeV and calculated the events at the far detector with the GLoBES \cite{globes} software using the following formula:
\begin{eqnarray} \nonumber
 N &=& \int \int \Phi(E_t) P_{\alpha \beta}(x, E_t) R(E_t, E_m) \sigma(E_t) M(E_t) \epsilon(E_t) \\ \nonumber && dE_t dE_m 
\end{eqnarray}
where $\Phi$ is the neutrino flux $E_t$ is the true energy, $E_m$ is the reconstructed energy, $P_{\alpha \beta}$ is neutrino oscillation probability for $\nu_\alpha \rightarrow \nu_\beta$ with $\alpha$, $\beta =$ e, $\mu$ and $\tau$, $x$ are the neutrino oscillation parameters, $R$ is the energy resolution function, $\sigma$ is the cross section, $M(E_{t})$ is the detector mass and $\epsilon(E_{t})$ is the efficiency factor. To calculate $N$ we used the following information: (i) both the neutrino and antineutrino fluxes are obtained from the P2O collaboration, (ii) the effective mass is taken from slide 8 of Ref. \cite{p2o_talk_zaborov}, (iii) the energy resolution is taken to be Gaussian with 30\% width as given in slide 24 of \cite{p2o_talk_zaborov}. Finally we implemented energy dependent efficiencies to match with the event spectrum in normal hierarchy as presented in slide 9 of \cite{p2o_talk_zaborov}. We have given in the left panel of Fig. \ref{events}. Then we generate events for IH taking the same specification which we used for normal hierarchy and this perfectly matches with the event spectrum as given in slide 10 of Ref. \cite{p2o_talk_zaborov}. This we have given in the right panel of Fig. \ref{events}. The choice of neutrino oscillation parameters are same as in slide 11 of \cite{p2o_talk_zaborov}. These events corresponds to three years running of the P2O experiment in the neutrino mode corresponding ot $12 \times 10^{20}$ proton on target\footnote{Note our method of calculating events is simplified and does not match-up with the sophistication that the analysis of the experimental collaboration would have. But nevertheless with this calculation of events, we successfully reproduce the sensitivity results of P2O.}. We have presented the energy dependence of the fluxes in Fig. \ref{flux}. The left (right) panel is for neutrinos (antineutrinos).  In the panels we have showed all the four components of the neutrino/antineutrino fluxes. The muon neutrino/antineutrino fluxes mainly play the major role in the sensitivity of P2O.
After matching the events our next aim is to match the $\chi^2$. But before that lets discuss the event topologies of the P2O experiment. 

The two main event topologies in ORCA are track and cascade. The tracks are almost always produced by $\nu_\mu$ charged current interactions while cascades come from $\nu_e$ charged current interactions, $\nu_\tau$ charged current interactions as well as neutral current interactions of neutrinos. In our simulations we have simulated track and cascade events for both neutrino and antineutrino channels by suitably modifying GLoBES. A brief discussion on the backgrounds follows. Since the tracks come from $\nu_\mu$ charged current interactions alone, the only backgrounds for the disappearance channel come from neutral current backgrounds\footnote{Note that though NC background is shower like, some of them can be confused with muon track events and is hence one of the indispensable backgrounds of the muon track events. However, this background does not play much role in the analysis of track events.} and tau backgrounds. For appearance channel, backgrounds come from $\nu_\tau$ charged current interactions and neutral current interactions as well as mis-identification\footnote{We will refer to this background as mis-id throughout our paper.} of the muon events as electron events. The   numbers for tau and neutral current background which we will use in our analysis are same as given in Fig. \ref{events}.  


For the fit method, we use the following informations: (i) We have considered both electron (appearance) and muon (disappearance) events, (ii) apart from the backgrounds mentioned above we have taken additional 20\% mis-id background in the appearance channel. Note that when we implement the mis-id background for appearance channel, we subtract the same mis-id factor from the disappearance channel signal events. For systematics we follow slide 11 of Ref. \cite{p2o_talk_zaborov}. We have taken an overall normalization error of 5\% for both appearance and disappearance channel signal. The background systematic error for disappearance channel is calculated by adding the normalization error for tau (10\%) and normalization error for neutral current (5\%) in quadrature which  is 11\%. For appearance channel background we have taken 6\% error. For the neutrino oscillation parameters we have taken the values as given in slide 11 of Ref. \cite{p2o_talk_zaborov}. With this specification we reproduce the $\chi^2$ plots of \cite{p2o_talk_zaborov}. These plots are hierarchy sensitivity curve vs $\theta_{23}$, CP violation sensitivity curve vs $\dcp$, CP precsion curves vs $\dcp$ and $\dcp$ resolution curve vs running time as given in slide numbers 12, 13, 14 and 15 of \cite{p2o_talk_zaborov} respectively.
Following \cite{p2o_talk_zaborov}, we have not considered any systematic error to incorporate a shift in the energy scale. 
Note that ref. \cite{p2o_talk_zaborov} does not give any information on antineutrinos, therefore we take the exact same values of the efficiencies, background and systematics for the antineutrino analysis. 


In the following sections we will discuss the hierarchy, octant and CP sensitivity of the P2O experiment. As the present data hints the preference of NH over IH, we present our results for NH in detail in the next section. After that we will give the results for IH for some selective values of parameters.

\section{Results for NH}
\label{results}

In this section we discuss the capability of the P2O experiment to discover the unknowns in the three flavour standard neutrino oscillations, namely neutrino mass hierarchy, octant of the mixing angle $\theta_{23}$ and measurement of the Dirac type phase $\dcp$. While calculating $\chi^2$ we have taken both the electron and muon events. Our aim of this study will be to find out the optimal configuration of the P2O experiment to discover the above mentioned parameters by considering various combinations of antineutrino exposure, systematic errors and backgrounds. For systematic uncertainty, we only vary the error associated with appearance channel signal and the only background we vary is the mis-id in the appearance channel.  
In our $\chi^2$ analysis we have kept fixed the parameters $\theta_{12} = 33.4^\circ$, $\theta_{13} = 8.42^\circ$, $\Delta m^2_{21} = 7.53 \times 10^{-5}$ eV$^2$ and $\Delta m^2_{31} = 2.44 \times 10^{-3}$ eV$^2$ in both the true and test spectrum. These values are consistent with the global analysis of the world neutrino data \cite{Esteban:2018azc,deSalas:2017kay,Capozzi:2018ubv}.

\subsection{Hierarchy}

The hierarchy sensitivity of an experiment is calculated by taking the correct hierarchy in the true spectrum and the wrong hierarchy in the test spectrum. First we study the effect of antineutrino run on the hierarchy sensitivity of the P2O experiment.
\begin{figure*}[htb]
 \begin{center}
\includegraphics[width=0.32\textwidth]{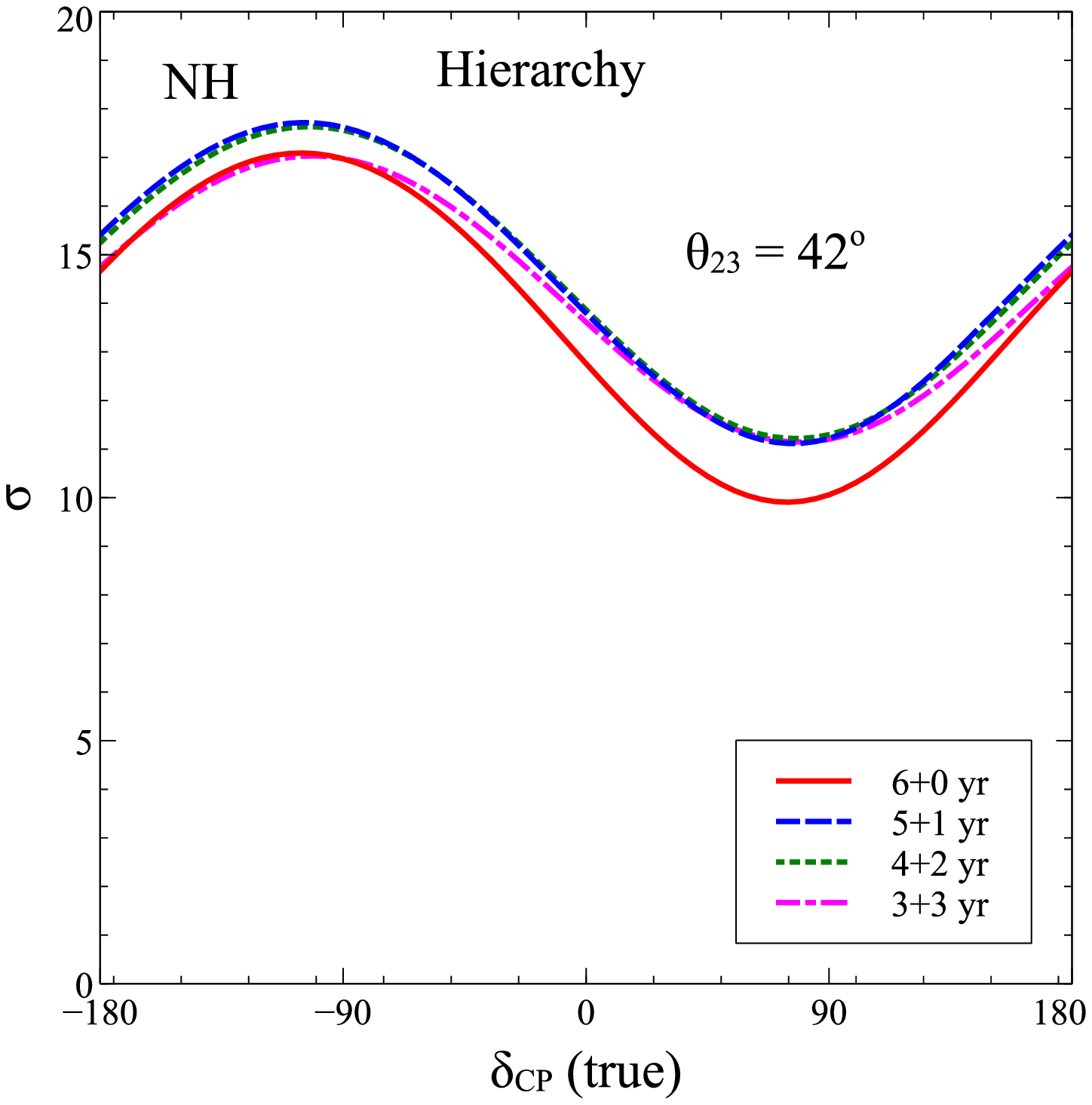} 
\includegraphics[width=0.32\textwidth]{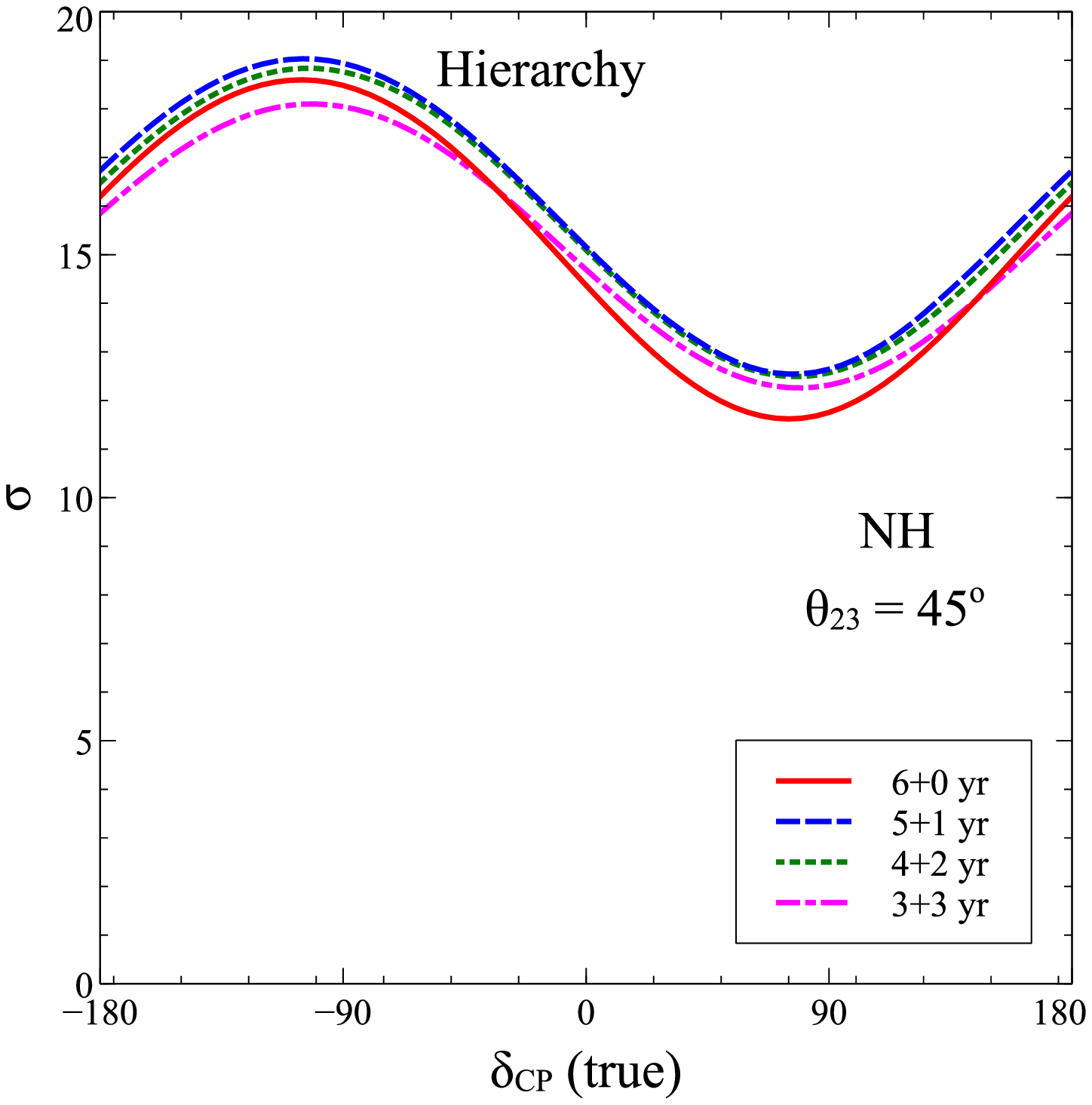}
\includegraphics[width=0.32\textwidth]{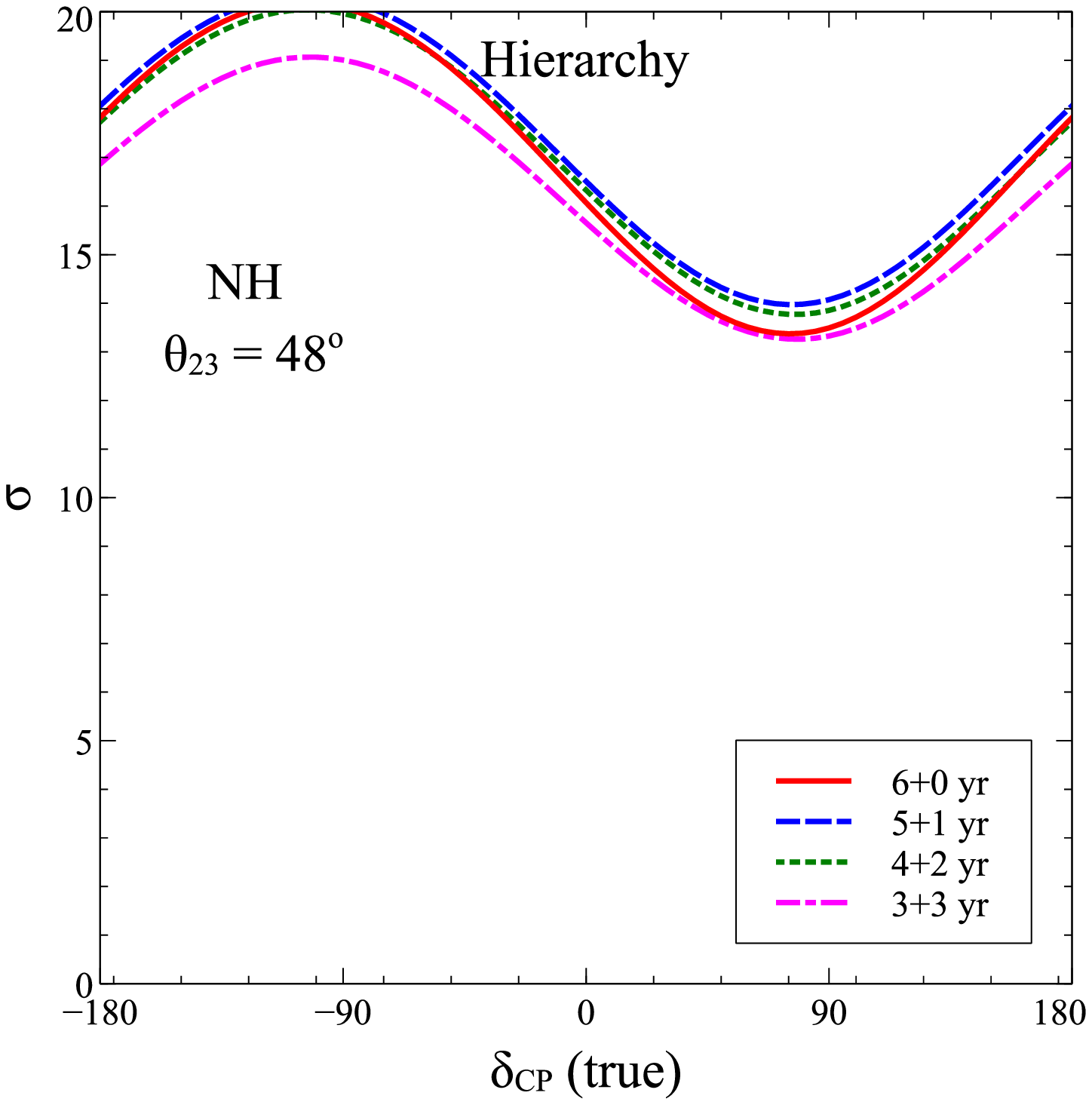} \\
\caption{Hierarchy sensitivity vs $\dcp$ (true) in NH for different combinations of neutrino and antineutrino ratio. The left/middle/right panel is for $\theta_{23}$(true)$ = 42^\circ$/$45^\circ$/$48^\circ$.}
\label{hier_anu}
 \end{center}
\end{figure*}
In Fig. \ref{hier_anu} we have plotted the hierarchy $\chi^2$ vs $\dcp$(true). The parameters 
$\theta_{23}$ and $\dcp$ have been marginalized in the test. 
The left/middle/right panel is for $\theta_{23}$ (true) value of $42^\circ$/$45^\circ$/ $48^\circ$. In these plots we have assumed a total run-time of six years and we have considered four different combinations of neutrino and antineutrino ratio. 

In all the three panels we see that the hierarchy sensitivity is maximum at $\dcp=-90^\circ$ and minimum at $\dcp=+90^\circ$. This is because of the well known hierarchy $\dcp$ degeneracy \cite{Prakash:2012az,Ghosh:2015ena}. In the case of P2O, this degeneracy is lifted because of the large matter effect. But still the probability in NH corresponding to $\dcp=-90^\circ$ ($+90^\circ$) is furthest (closest) to the probability in the IH. This is true for both neutrino and antineutrino. To understand the role of antineutrinos we need to understand the behavior of octant degeneracy. It was shown in \cite{Ghosh:2015tan,Prakash:2013dua} that the main role of the antineutrino run is to resolve the octant degeneracy to improve the CP and hierarchy sensitivity in the long-baseline experiments. Therefore, if the wrong hierarchy $\chi^2$ occurs with the wrong octant, then addition of antineutrinos can improve the hierarchy $\chi^2$ by removing the wrong octant solution. That is what is happening for $\theta_{23} = 42^\circ$ (left panel). For 6+0 years, the hierarchy $\chi^2$ occurs with the wrong octant and 5+1 years gives better sensitivity than 6+0 years. In this panel we also note that further addition of antineutrinos decreases the sensitivity. From this we understand that only one year of antineutrino is sufficient to remove the octant degeneracy and further addition of antineutrino only decreases the statistics due its low cross section and smaller flux as compared to the neutrinos. The situation is slightly different for $\theta_{23}=48^\circ$ (right panel). At this true value of $\theta_{23}$ the hierarchy $\chi^2$ appears with the right octant and thus antineutrino should not help much. That is why we see that the sensitivity is almost best for 6+0 years for $\dcp=-90^\circ$ (true). However the best sensitivity is obtained for 5+1 years. Now let us discuss for $\theta_{23}=45^\circ$ where there is no octant degeneracy (middle panel). Here we see that at $\dcp({\rm true})=-90^\circ$, 5+1 years give best sensitivity and with further addition of antineutrinos, sensitivity decreases. Whereas for $\dcp ({\rm true})=90^\circ$, 5+1 years gives the best sensitivity and the worst sensitivity comes for 3+3 years and 6+0 years.

From these three panels we see that the minimum hierarchy sensitivity for six year running of the experiment is always greater than $10 \sigma$, irrespective of the antineutrino exposure. 

\begin{figure}[htb]
 \begin{center}
\includegraphics[width=0.42\textwidth]{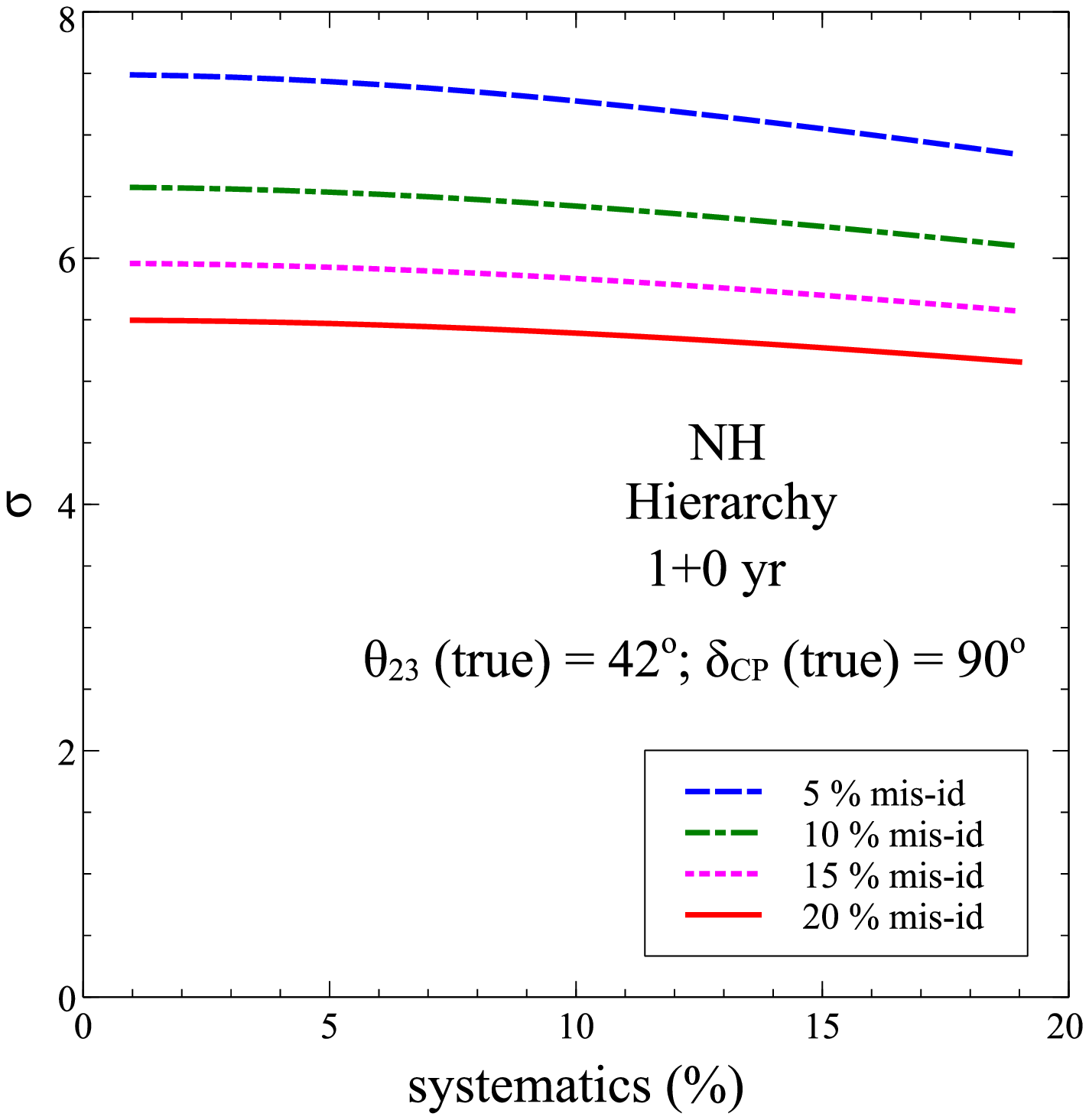} 
\caption{Effect of background and systematics in hierarchy sensitivity for NH. This is shown for $\theta_{23}$(true)$=42^\circ$ and $\dcp$(true)$=90^\circ$ as for this set of parameters the $\chi^2$ is minimum in NH. 
}
\label{hier_bg_sys}
 \end{center}
\end{figure}

Next we study the effect of background and systematics on hierarchy sensitivity. In Fig. \ref{hier_bg_sys}, we have presented the hierarchy $\chi^2$  vs systematic uncertainty for the four combinations of background. We present this for just one year of neutrino exposure of the P2O experiment and for $\dcp ({\rm true})=+90^\circ$ and $\theta_{23}({\rm true})=42^\circ$. As the $\chi^2$ is minimum for this value of $\dcp$ and NH true, this gives the most conservative configuration of the experiment. From this plot we understand that the sensitivity does not vary much with the systematic uncertainty. For all the four combinations of the background, hierarchy sensitivity falls only within $1 \sigma$ when systematic is varied from 1\% to 20\%. On the other hand we observe that the sensitivity depends greatly on the amount of background. For 5\% systematic error, the sensitivity goes from $5.5 \sigma$  to $7.5 \sigma$ when background decreases from 
20\% to 5\%. However, it is very important to note from this plot that whatever be the systematic error or the background, P2O has hierarchy sensitivity always more than $5 \sigma$, even with one year neutrino run. This can be attributed to two facts: (i) huge matter effect enable P2O to have a very large sensitivity and (ii) the baseline for P2O is close to the bi-magic baseline \cite{Dighe:2010js,Raut:2009jj} for which the hierarchy sensitivity gets enhanced at a particular energy.

\subsection{Octant}

Octant sensitivity of an experiment is calculated by taking the correct octant in the true spectrum and wrong octant in the test spectrum. We first study the effect of antineutrinos in the determination of the octant.
\begin{figure*}[htb]
 \begin{center}
\includegraphics[width=0.32\textwidth]{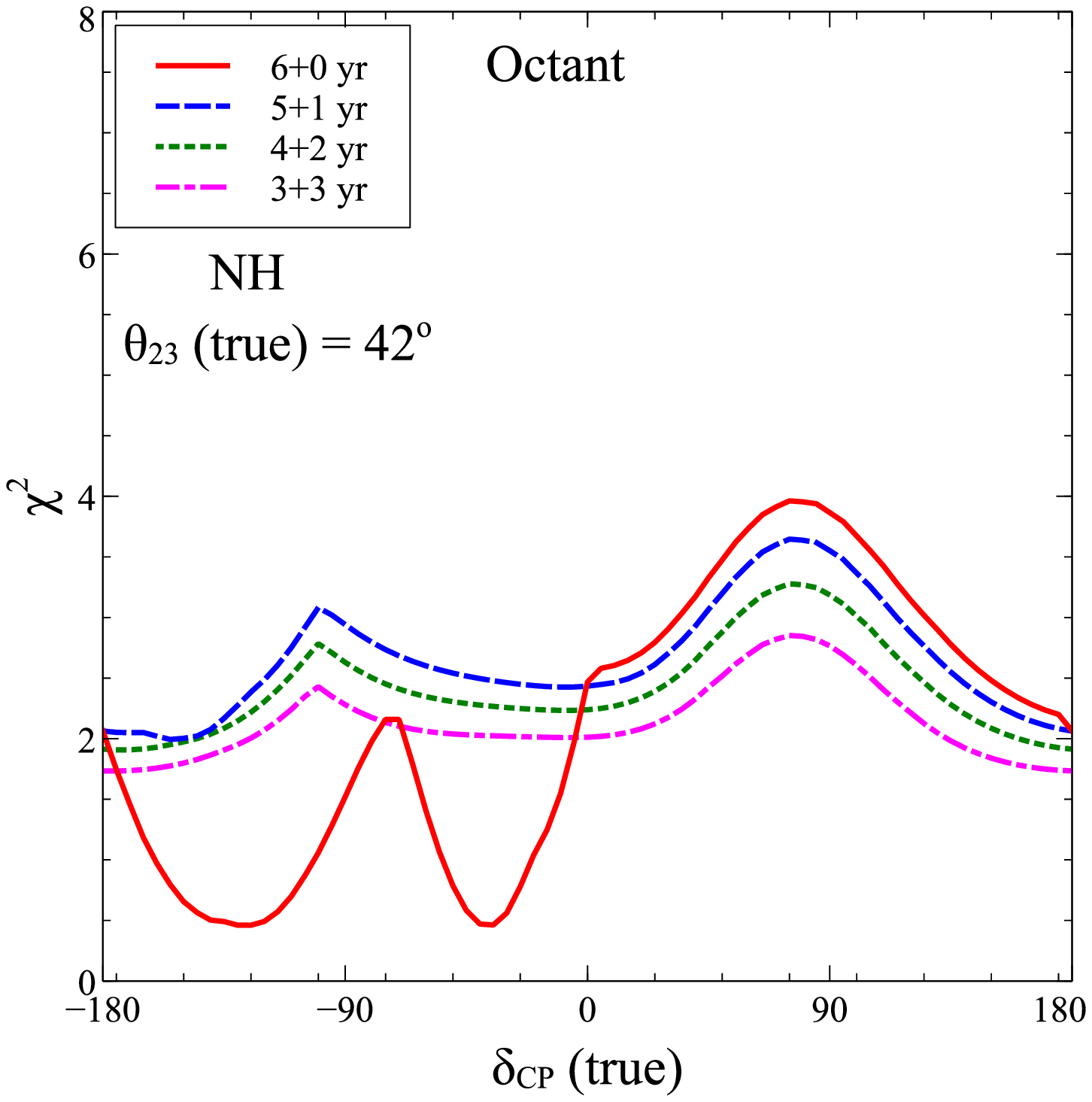} 
\includegraphics[width=0.32\textwidth]{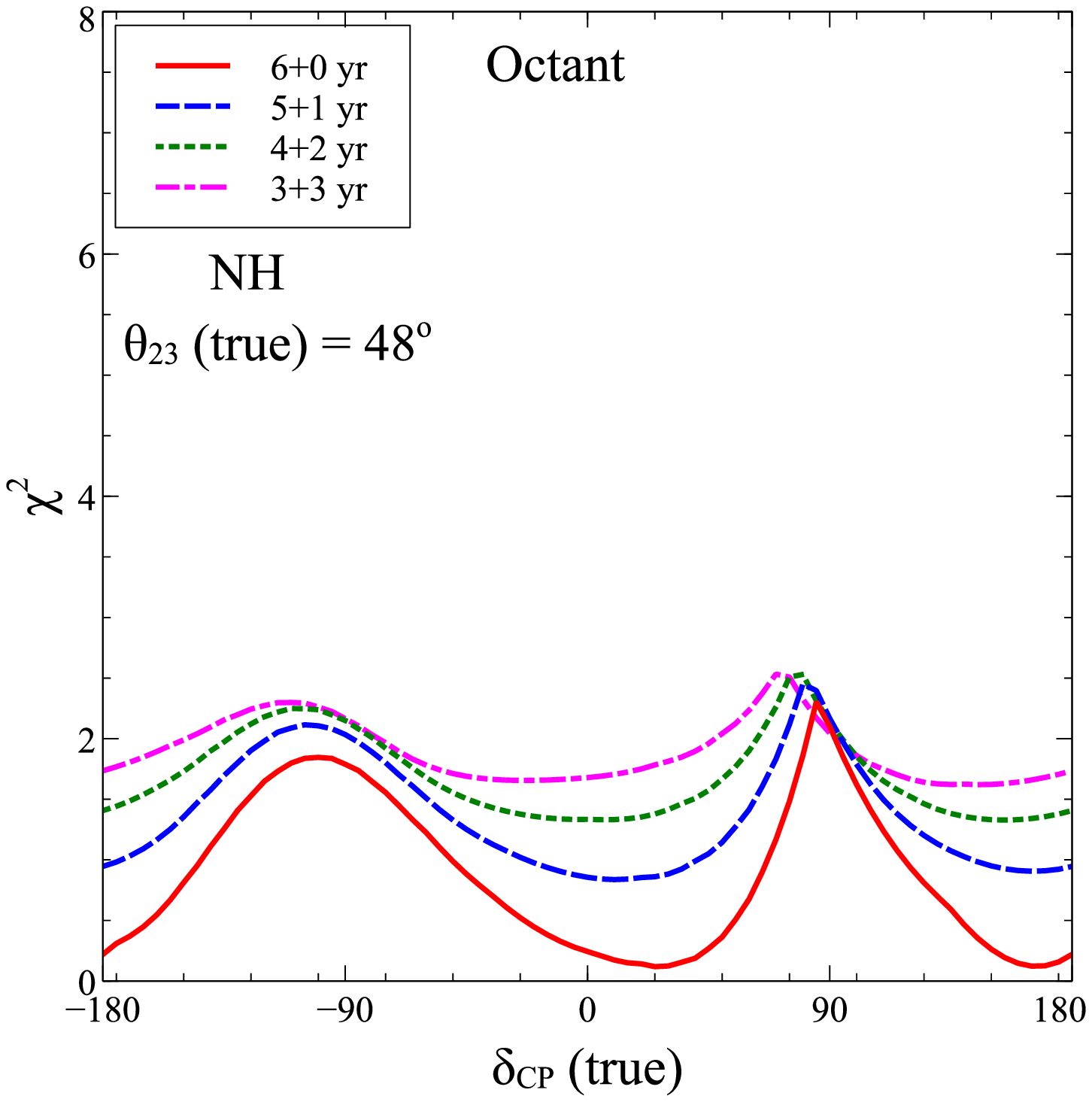}
\includegraphics[width=0.32\textwidth]{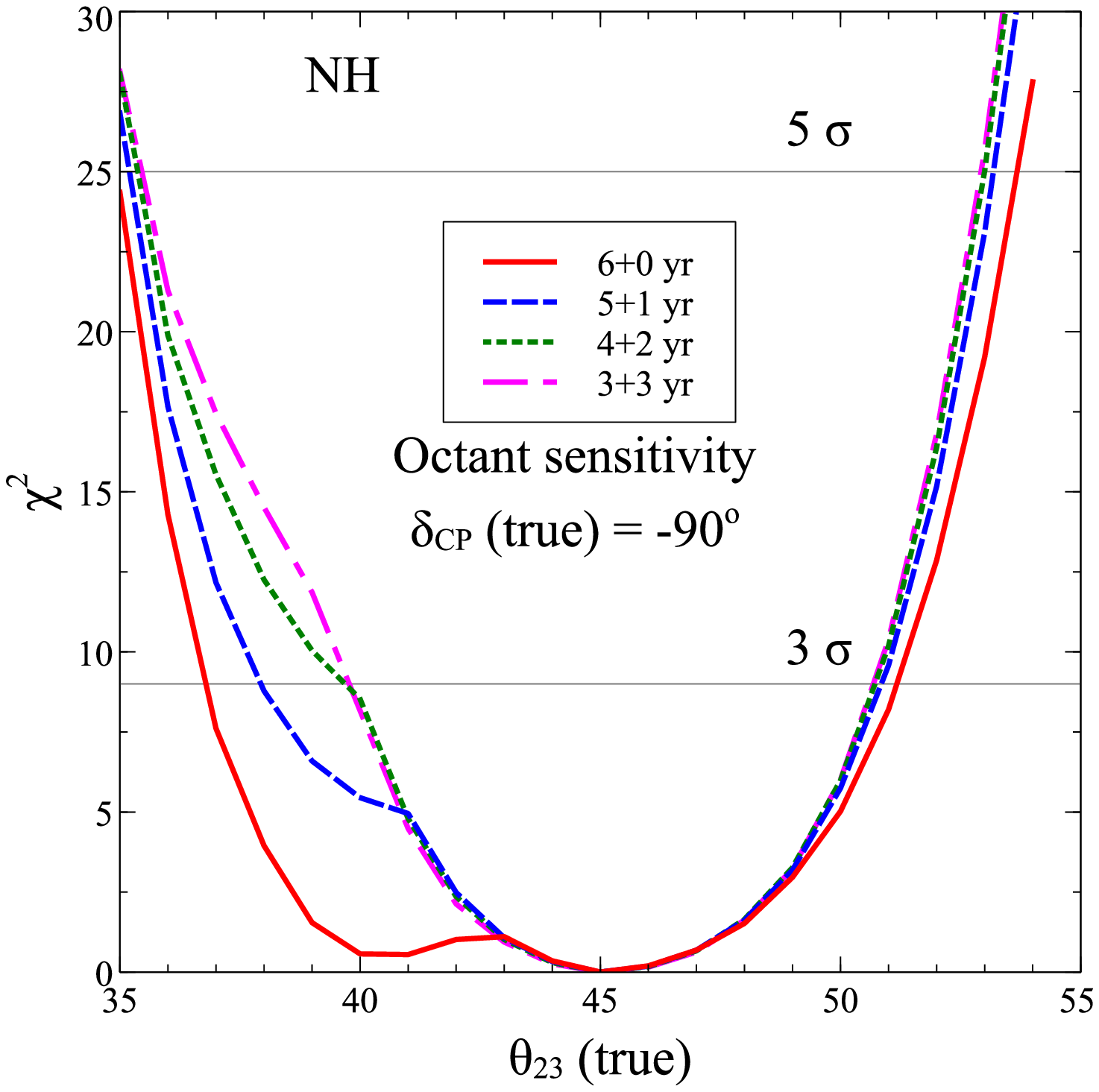} \\
\caption{Octant sensitivity in NH for different combinations of neutrino and antineutrino ratio. Octant $\chi^2$ vs $\dcp$(true) is presented in the left and middle panel for $\theta_{23}=42^\circ$ and $48^\circ$ respectively. Octant $\chi^2$ vs $\theta_{23}$(true) is presented in the right panle for $\dcp=-90^\circ$.}
\label{oct_anu}
 \end{center}
\end{figure*}
In Fig. \ref{oct_anu}, we have presented the octant sensitivity for different combinations of neutrino and antineutrino ratio, taking the total run-time of six years. In the left and middle panel we have presented the results as a function of $\dcp$ (true) and in the right panel we have given our results as a function of $\theta_{23}$ (true).The left and middle panel are for $\theta_{23}$ in the lower octant ($\theta_{23}=42^\circ$) and higher octant ($\theta_{23}=48^\circ$), respectively. In the right panel, true value of $\dcp=-90^\circ$ which is the present best-fit value of this parameter. In all the panels $\theta_{23}$ has been marginalized in the opposite octant and $\dcp$ has been marginalized in its full range. Hierarchy is also marginalized.

To understand the role of antineutrinos in resolving octant degeneracy, first we have to understand how octant degeneracy occurs in neutrinos and antineutrinos. It has been shown in \cite{Agarwalla:2013ju,Ghosh:2015ena} that for the neutrino channel octant degeneracy occurs at ($-180^\circ < \dcp < 0^\circ$, LO) with ($0^\circ < \dcp < 180^\circ$, HO) and for the antineutrino channel octant degeneracy occurs at ($0^\circ < \dcp <180^\circ$, LO) with ($-180^\circ < \dcp < 0^\circ$, HO). This is the reason why a balanced antineutrino run is always important to have significant octant sensitivity. The above discussion explains why for $42^\circ$ octant sensitivity is maximum for 6+0 years for $0^\circ < \dcp < +180^\circ$ and minimum for $-180^\circ < \dcp < 0^\circ$ (left panel). With the addition of an antineutrino run the sensitivity for $\dcp=+90^\circ$ decreases and sensitivity for $\dcp=-90^\circ$ increases. From the plot we also realise that the 5+1 years is an optimal choice of exposure. For this exposure, the octant sensitivity of $\chi^2=3.5 (4)$ can be obtained for $\dcp = -90^\circ$ ($+90^\circ$). For $\theta_{23}=48^\circ$ we notice that the sensitivity is maximum for 3+3 years at $\dcp=-90^\circ$ and minimum for 6+0 years. Whereas for $\dcp=+90^\circ$ all the combinations of the antineutrino run give almost equal sensitivity. Note that the sensitivity in the higher octant is smaller than the sensitivity in the lower octant. This is due to the fact that for the higher octant, the denominator in the $\chi^2$ is higher than the denominator in the $\chi^2$ for lower octant. Here the octant sensitivity is around $\chi^2 = 2.5 $ for both $\dcp = -90^\circ$ and $+90^\circ$. The right panel shows octant sensitivity for different values of $\theta_{23}$ for $\dcp=-90^\circ$. From the plot we see that in the lower octant, 6+0 years gives the worst octant sensitivity and in the region $41^\circ < \theta_{23} < 45^\circ$, 5+1 years give the best octant sensitivity. For higher octant all the four combination give almost equal sensitivity. From this plot we conclude that with the current background and systematics, P2O can resolve octant at $3 \sigma$ for $\theta_{23}$ values except $41^\circ (39^\circ) < \theta_{23} < 51^\circ$ for 3+3 years (5+1 years) run.

\begin{figure}[htb]
\begin{center}
\includegraphics[width=0.42\textwidth]{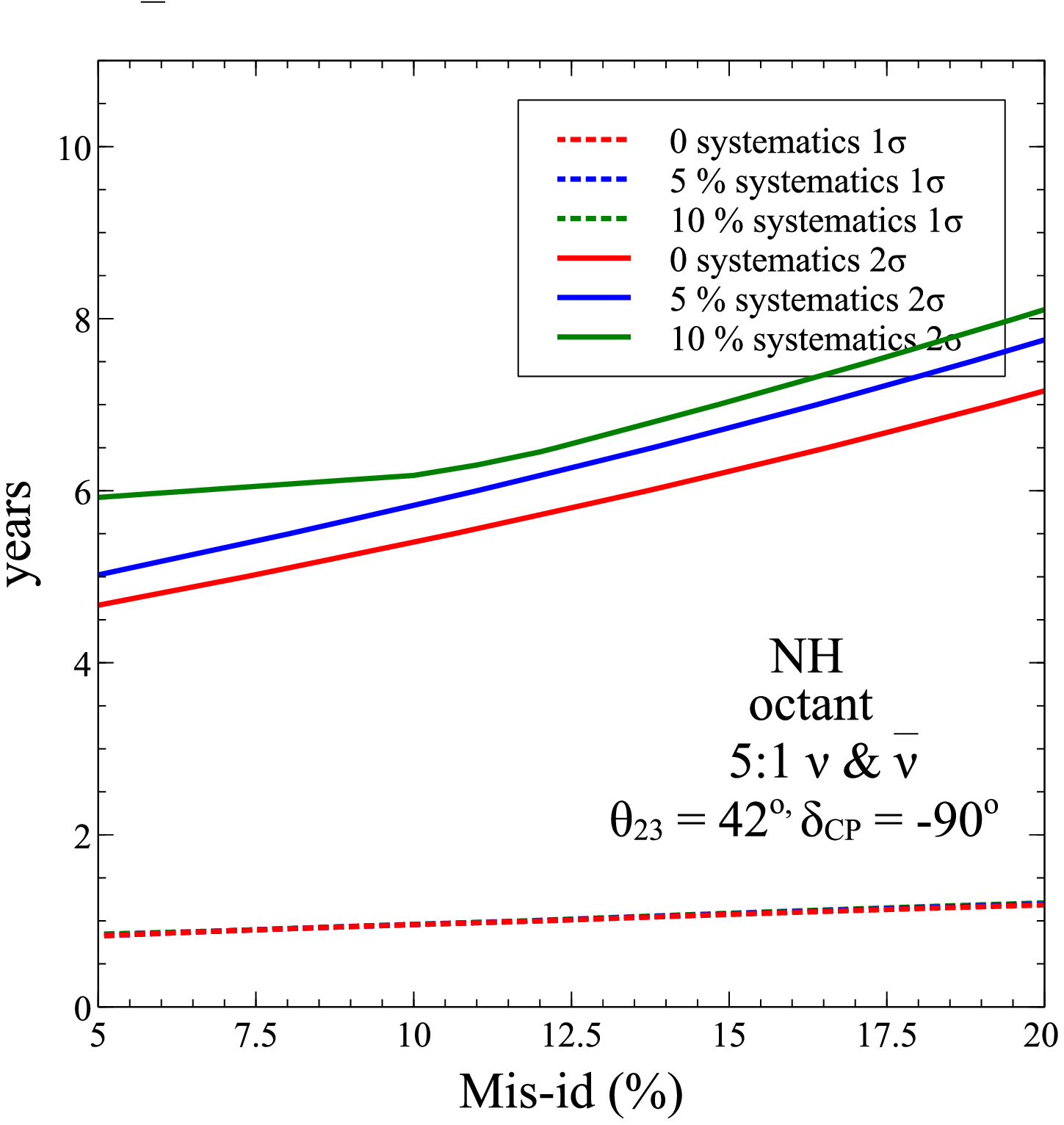}
\end{center}
\caption{Effect of background and systematics in octant sensitivity for NH in terms of runtime required to acheieve a sensitivity of a certain confidence level. This is shown for $\theta_{23}$(true)$=42^\circ$ and $\dcp$(true)$=-90^\circ$ as for this set of parameters are close to the present best-fit.  
}
\label{oct_bg_sys}
\end{figure}

\begin{figure*}[htb]
 \begin{center}
\includegraphics[width=0.32\textwidth]{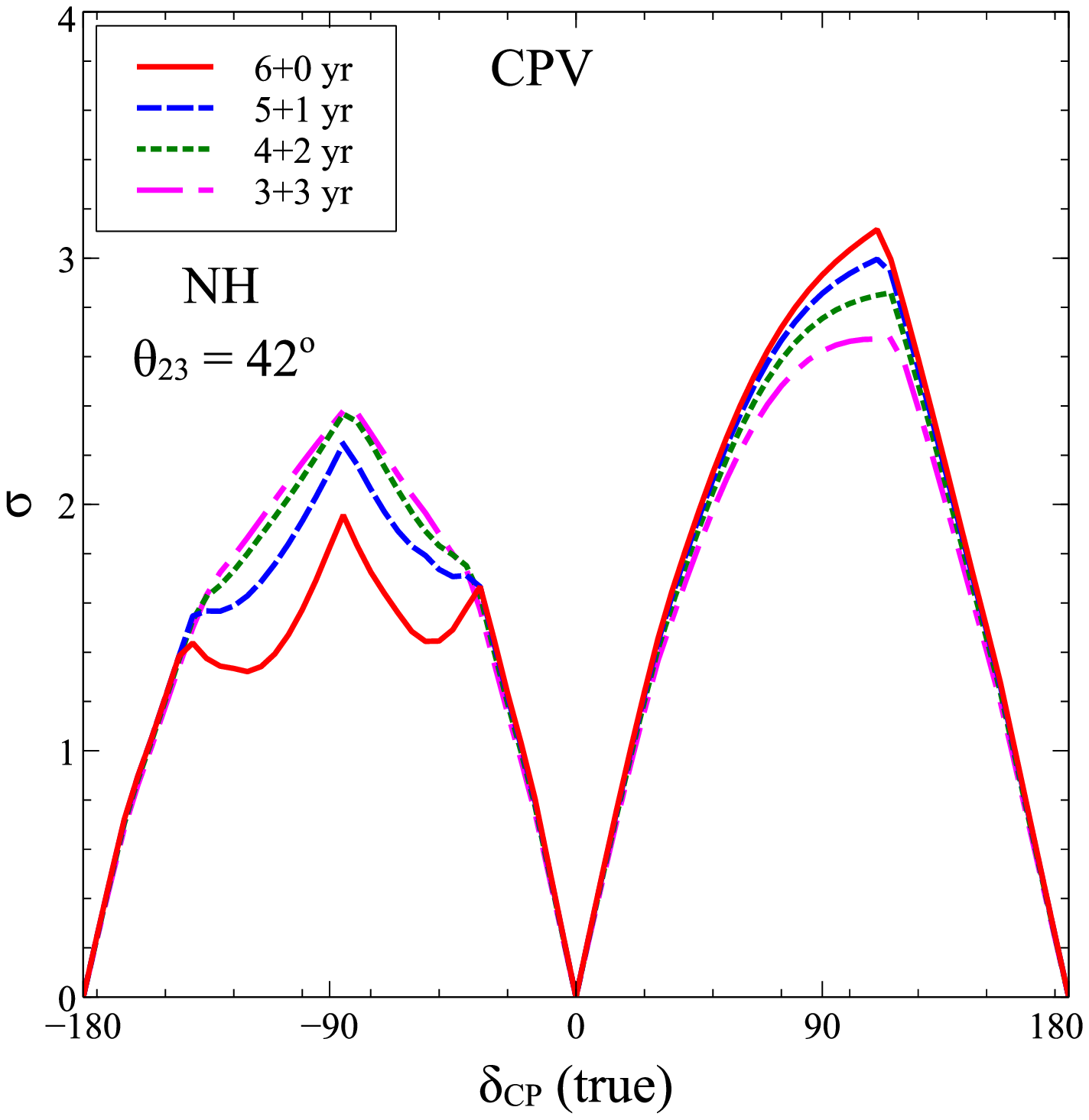} 
\includegraphics[width=0.32\textwidth]{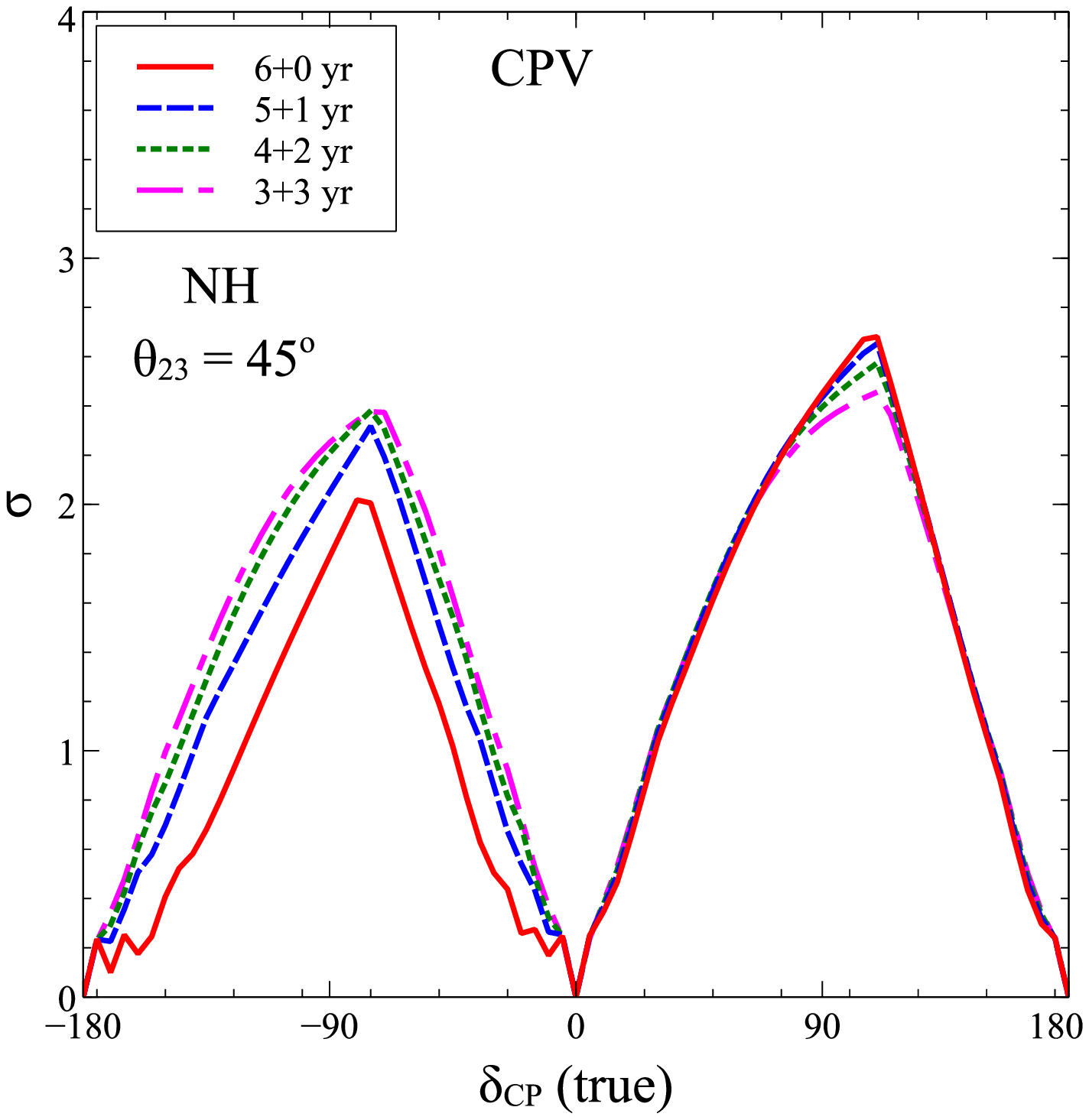}
\includegraphics[width=0.32\textwidth]{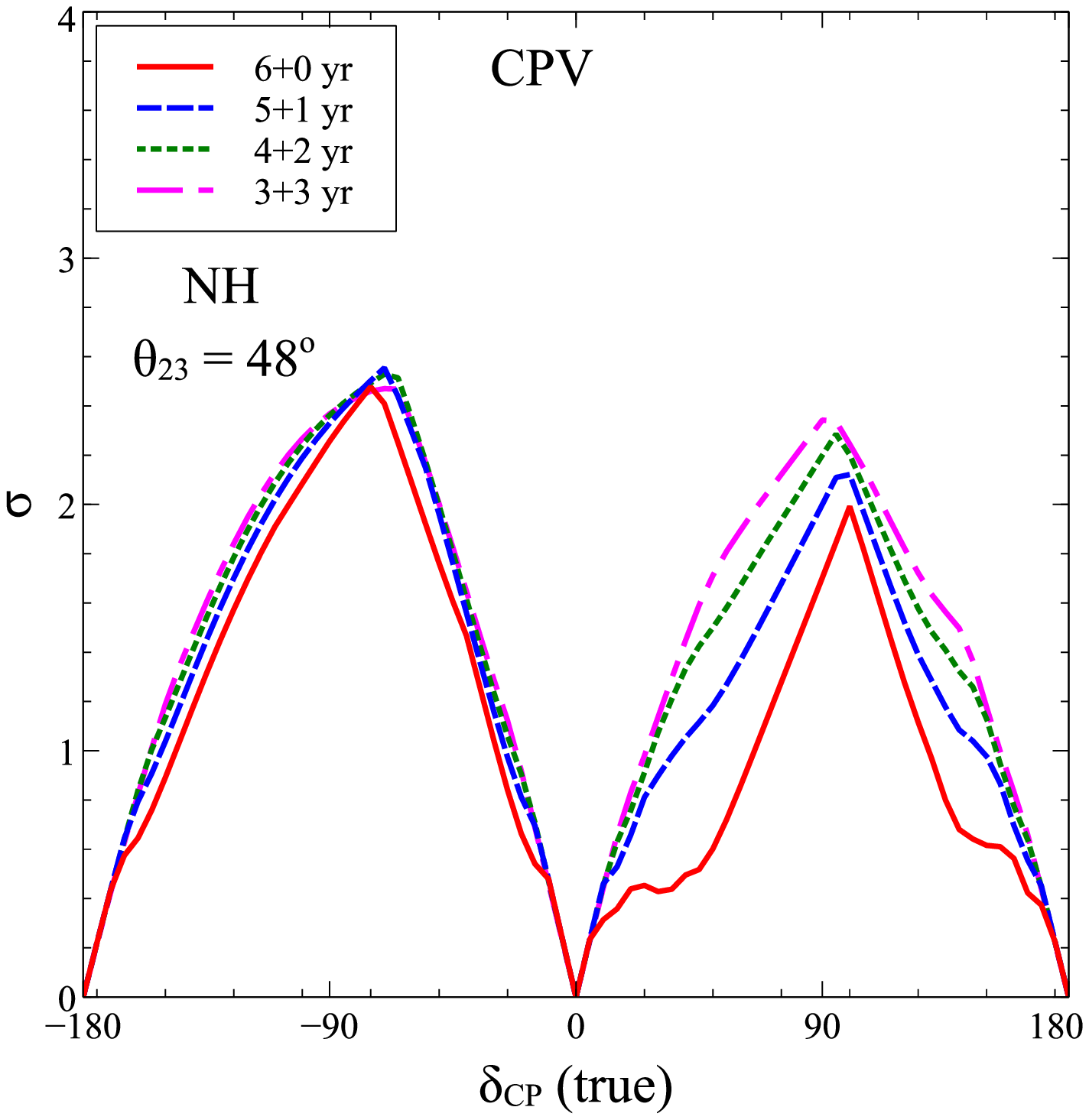} \\
\caption{CPV discovery sensitivity vs $\dcp({\rm true})$ for different combinations for neutrino and antineutrino in NH. The left/middle/right panel is for $\theta_{23} = 42^\circ$/$45^\circ$/$48^\circ$.}
\label{cpv_anu}
 \end{center}
\end{figure*}

\begin{figure}[htb]
\begin{center}
\includegraphics[width=0.42\textwidth]{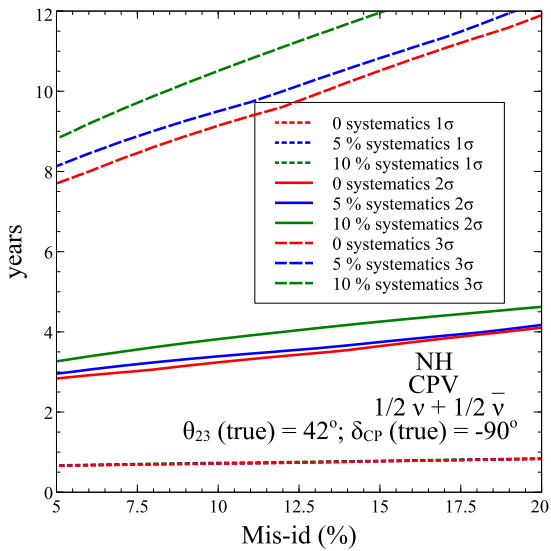}
\end{center}
\caption{Effect of background and systematics for CPV discovery in NH. in terms of runtime required to acheieve a sensitivity of a certain confidence level. This is shown for $\theta_{23}$(true)$=42^\circ$ and $\dcp$(true)$=-90^\circ$ as for this set of parameters are close to the present best-fit.
}
\label{cpv_bg_sys}
\end{figure}

\begin{figure}[htb]
\begin{center}
\includegraphics[width=0.42\textwidth]{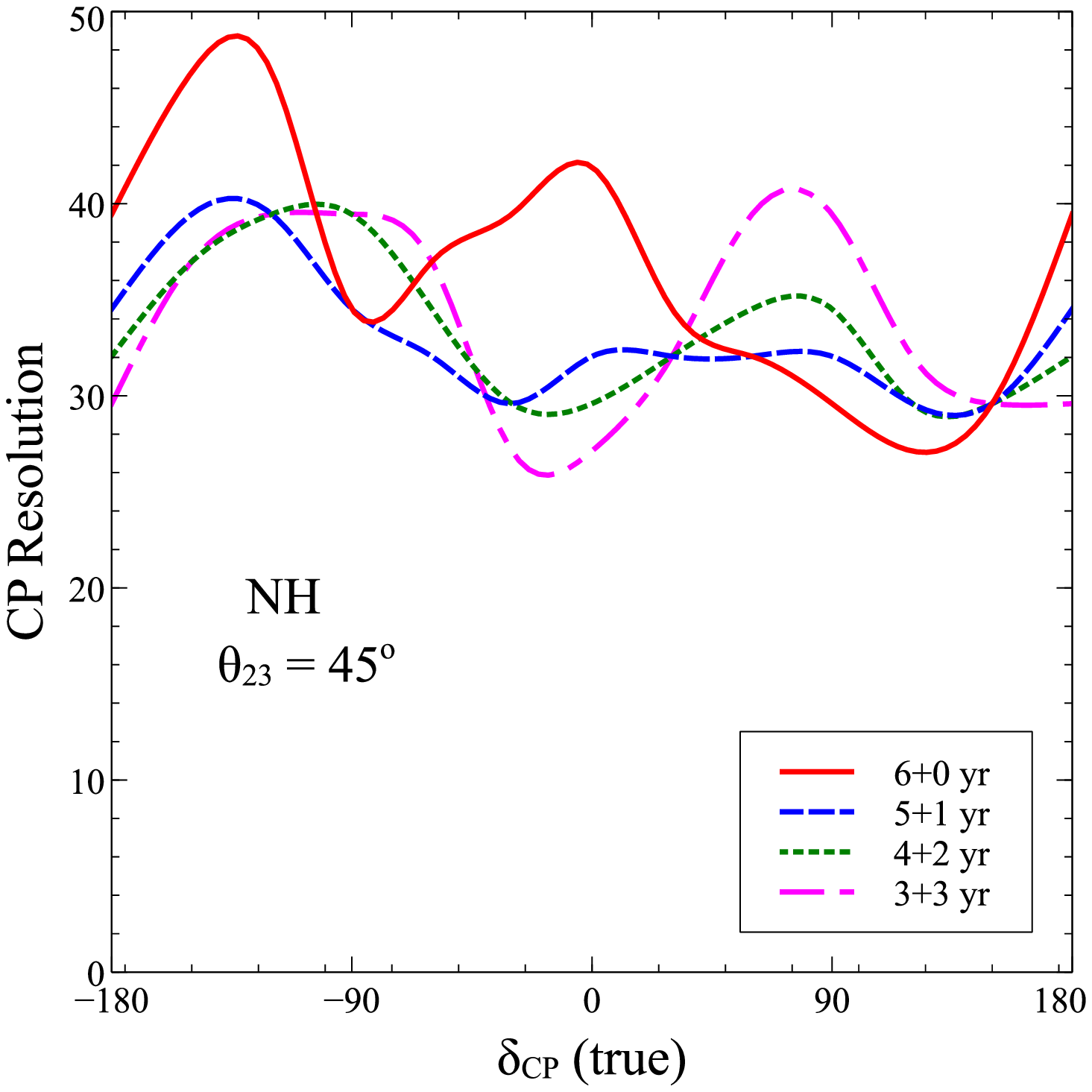}
\end{center}
\caption{CP resolution sensitivity in NH. This is shown for $\theta_{23}$(true)$=45^\circ$ and taking four different combinations of neutrino-antineutrino ratio.
}
\label{cpp_anu}
\end{figure}

From the above discussion we realize that 5+1 years would be the best combination for P2O to determine the octant sensitivity for both LO and HO. But even for this best combination, a maximum $\chi^2$ of 4 can be achieved at the present best-fit of $\dcp$ and $\theta_{23}$ which is $-90^\circ$ and $42^\circ \&$ 48$^\circ$. Next we study the effect of background and systematics in octant sensitivity with the neutrino to antineutrino ratio of 5:1.

In Fig. \ref{oct_bg_sys}, we have presented the run-time to achieve a sensitivity of a certain confidence level as a function of background for different combination of systematic uncertainties. we present this for $\theta_{23}=42^\circ$ and $\dcp=-90^\circ$. From the figure we see that there is no effect of background and systematics to achieve $1 \sigma$ sensitivity as the required run-time remains almost fixed to almost 1 year for all the values of systematics and background. To achieve $2 \sigma$ sensitivity the run-time keeps increasing as we increase the systematics for a fixed value of background. For the most optimistic case (i.e., 0\% systematic error), the required run-time increases from 4.5 years to 7 years to achieve a $2\sigma$ octant sensitivity when the background increases from 5\% to 20\%.

\subsection{CP}

\begin{figure*}[htb]
\includegraphics[width=0.45\textwidth]{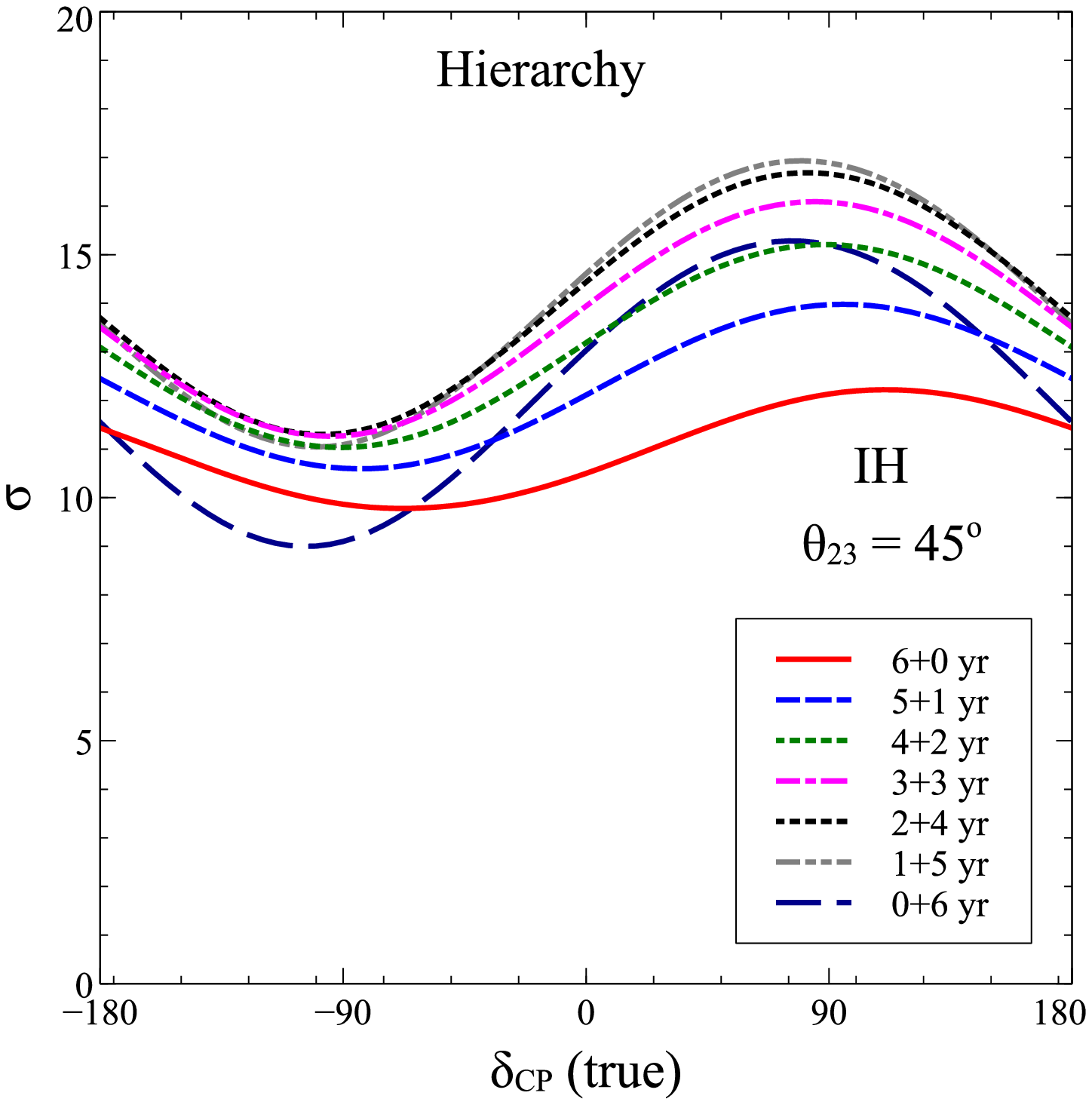} 
\hspace{0.5 in}
\includegraphics[width=0.45\textwidth]{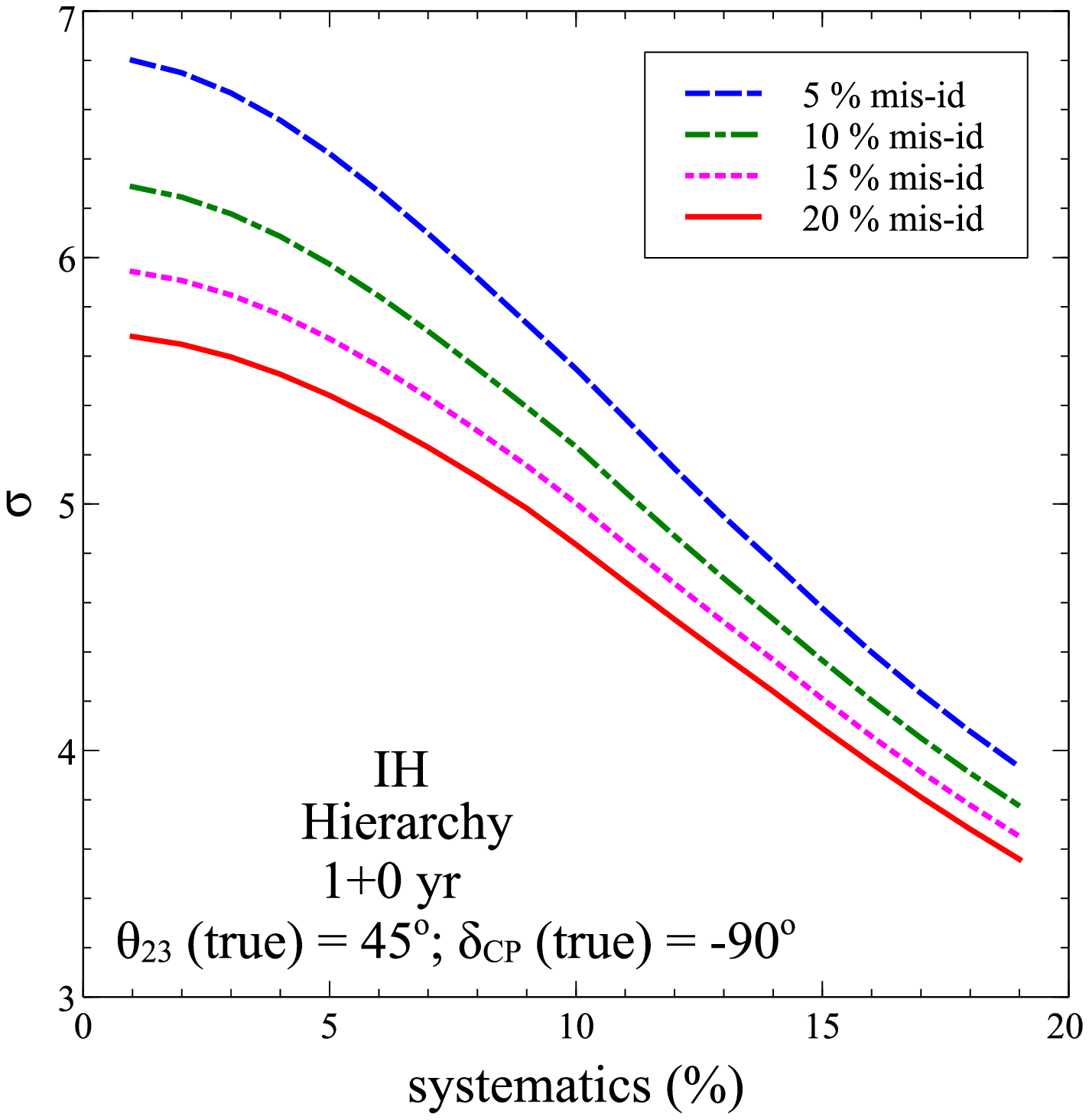}
\caption{Hierarchy sensitivity for different combinations for neutrino and antineutrino in IH is shown in the left panel for $\theta_{23}=45^\circ$. The right panel shows the impact of systematics and backgrounds on the hierarchy sensitivity in IH for $\theta_{23}=45^\circ$ and $\dcp=-90^\circ$.}
\label{hier_IH}
\end{figure*}
First we discuss the CP violation (CPV) discovery potential of the P2O experiment. CP violation discovery potential is defined as the capability to distinguish a value of
$\dcp$ from the CP conserving values of $0^\circ$ and $180^\circ$. To study the role of antineutrinos, in Fig. \ref{cpv_anu}, we have plotted sensitivities for four different combinations of neutrino to antineutrino ratio. The left/middle/right panel is for $\theta_{23}=42^\circ/45^\circ/48^\circ$. In these figures we have marginalized over $\theta_{23}$ and hierarchy in the test. The effect of octant degeneracy can be seen for ($\theta_{23}$, $\dcp$) combination of ($42 ^\circ$, $-90^\circ$) and ($48 ^\circ$, $+90^\circ$) from the shape of the curves. This is because for the pure neutrino run (i.e., 6+0 years) the CPV $\chi^2$ occurs in the wrong octant for ($42 ^\circ$, $-90^\circ$) and ($48 ^\circ$, $+90^\circ$). Therefore the addition of an antineutrino run is supposed to help in the sensitivity by removing the octant degeneracy. From the panel we understand that this is exactly what is happening as the sensitivity keeps improving with the addition of antineutrino data. However from the shape of the curves it is also clear that even with the addition of antineutrino runs, the octant degeneracy is not completely removed. For the other two combinations i.e., ($42^\circ$, $+90^\circ$) and ($48^\circ$, $=-90^\circ$) we note that 6+0 years and 5+1 years give the best sensitivity, respectively. The middle panel shows the CPV sensitivity without any octant degeneracy. From this panel we note that addition of antineutrino run helps in improving the CPV sensitivity. For $\dcp=-90^\circ$, 3+3 years give best sensitivity and for $\dcp=+90^\circ$, 6+0 years provide the best sensitivity. From the discussion we can also conclude that the antineutrino run is important for CPV discovery and 5+1 years will be the best option for P2O. It is also important to note that irrespective of the combination of the antineutrino run, the CPV discovery is always less than  $3\sigma$ for all the three combinations of $\theta_{23}$.

\begin{figure*}[htb]
\includegraphics[width=0.45\textwidth]{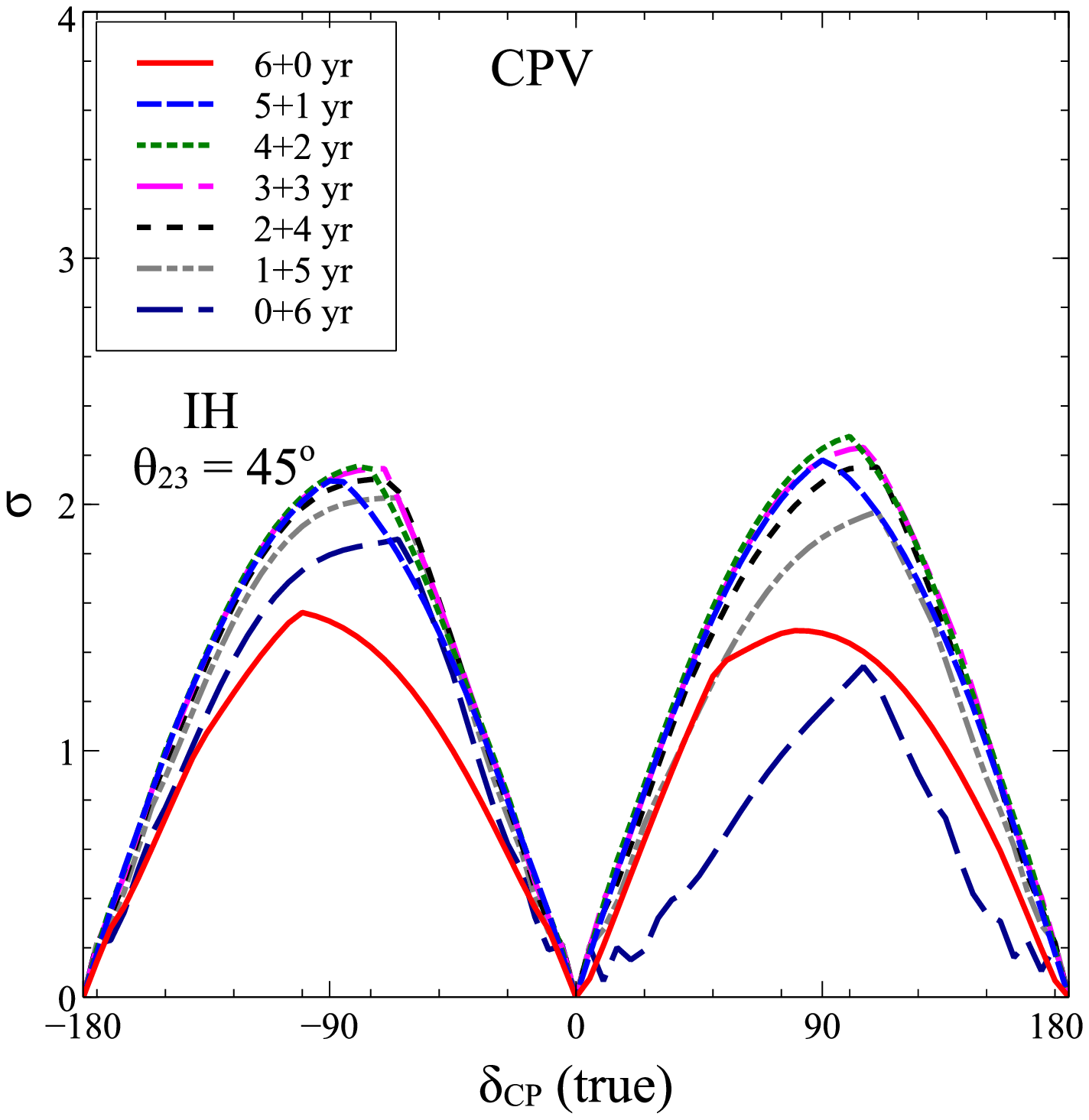} 
\hspace{0.5 in}
\includegraphics[width=0.45\textwidth]{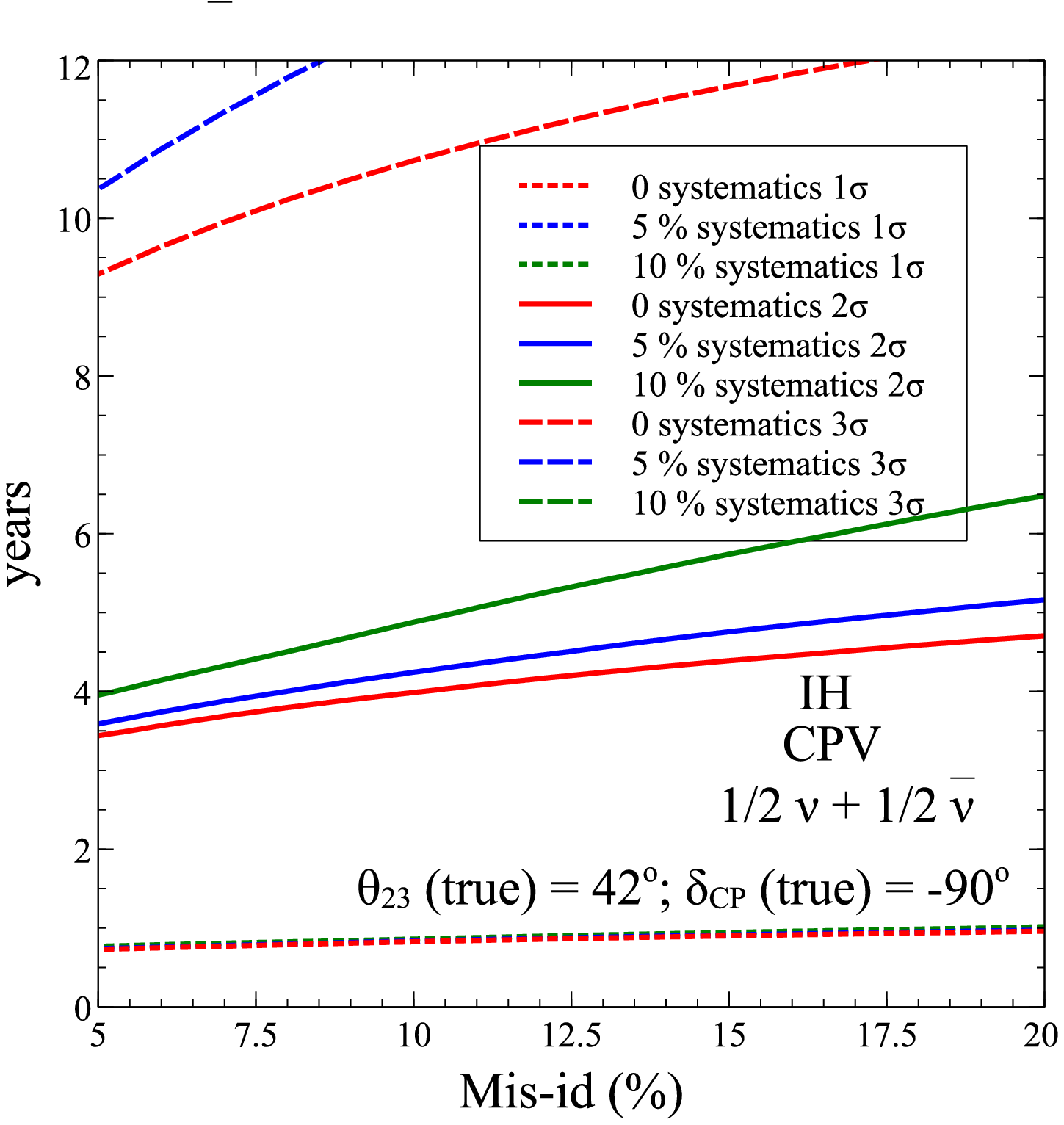}
\caption{The left panel shows the CPV sensitivity as a function of $\dcp$(true) in IH for $\theta_{23}=45^\circ$. The right panel shows the impact of systematics and backgrounds on the CPV sensitivity in IH in terms of runtime required to acheieve a sensitivity of a certain confidence level. This is for $\theta_{23}=45^\circ$ and $\dcp=-90^\circ$.}
\label{cpv_IH}
\end{figure*}

We next focus on the effect of background and systematics in the CPV sensitivity of P2O.
In Fig. \ref{cpv_bg_sys} we have plotted the time required to achieve a CPV sensitivity of a certain confidence level vs the background for different combination of the systematic uncertainty. We have done this for $\theta_{23}= 42^\circ $ and $\dcp = -90^\circ$. The curves are for equal ratios of neutrinos and antineutrinos. From the plot we see that to achieve $1 \sigma$ sensitivity, P2O requires around 1 year of running, irrespective of the value of systematic and background. The effect of systematic and background comes into the picture to achieve a sensitivity at a higher confidence level. To achieve a $2 \sigma$ CPV discovery, P2O will require around 3.5 years if the background is 5\% and 5.5 years if the background is 20\%. This is true irrespective of the systematic error. From the figure we also note that the systematic uncertainty starts to play some role when P2O tries to achieve a CPV sensitivity of $3 \sigma$ as the curves for different systematic errors tends to separate from each other. Here we observe that the required run-time of P2O is more than 10 years to achieve a $3 \sigma$ CPV discovery if the background is larger than $7.5\%$.

At this point we want to stress the fact that although P2O will achieve a $5 \sigma$ hierarchy sensitivity in less than 1 year, it can only achieve a $2 \sigma$ CPV and octant sensitivity in 6 years of run-time. As we mentioned earlier the hierarchy sensitivity at this baseline is enhanced due to larger matter effect and bi-magic property. The lack of octant and CP sensitivity stems from the poor energy resolution and particle identification. One solution can be optimization using more dense instrumentation of the ORCA detector which can improve the energy resolution and particle identification.

Let us now turn to how precisely P2O will be able to measure a value of $\dcp$.
To study this, in Fig. \ref{cpp_anu} we have plotted the CP resolution for each value of true $\dcp$ for $\theta_{23} ({\rm true}) = 45^\circ$. We define CP resolution by 0.5 $\times$ allowed range of $\dcp$ values at 1 $\sigma/360$. In this figure we have marginalized over $\theta_{23}$ and hierarchy in the test. This we have presented for four combination of antineutrino runs. From the curves we see that the CP precision is best around $0^\circ$ and $180^\circ$ and worst around $\pm 90^\circ$. Apart from these we can also notice a few local mininma which are more prominent in the case of pure neutrino run and getting removed when antineutrino run is added. This is due to the parameter degeneracies which are present in the pure neutrino run and getting removed when antineutrino run is added. From this figure we understand that the best CP resolution is achieved for the 5+1 years combination. For this combination a $\dcp$ can be measured within 34/32/32 degrees uncertainty for $\dcp({\rm true}) = -90^\circ/0^\circ/+90^\circ$. 

\section{Discussion for Sensitivity in IH}
\label{inverted}

In this section we discuss the sensitivity of the P2O experiment in IH. In Fig.~\ref{hier_IH} we present the hierarchy sensitivity for $\theta_{23}=45^{\circ}$. In the left panel, we have given the sensitivity vs $\dcp({\rm true})$ for a total run-time of six years. Like the case of NH, we have divided the run-time for different combinations of neutrino and anti-neutrino ratio. Apart from the four combinations given for the NH case, we have also put 2+4 years, 1+5 years and 0+6 years run-times. We have included the dominating anti-neturino run-times for IH because matter effect is reversed for the inverted hierarchy case. Here we can see that indeed anti-neutrino dominating run-times  2+4 years and 1+5 years are better, however, for the case of 0+6 years the sensitivity is again lower. This is because the statistics in the antineutrino channel is smaller compared to the statistics in the neutrino channel.  
In the right panel of Fig. \ref{hier_IH}, we have shown the variation of the hierarchy sensitivity with respect to background and systematics for a run-time of 1+0 year. We have done this for $\dcp= -90^\circ$ for which the sensitivity in IH is minimum. This will give the most conservative estimate. From this we note that the effect of systematics is stronger in IH as compared to NH. 
This plot shows that to achieve a $5 \sigma$ sensitivity, the systematic error can be allowed to increase from 9\% to 13\% if the background decreases from 20\% to 5\%.

Next let us discuss the octant sensitivity of the P2O experiment in IH. The octant sensitivity of the P2O experiment for IH is very poor. We have checked that for a total run-time of six years, the octant sensitivity in pure neutrino run (i.e., 6+0 years) is negligible for both $\theta_{23} = 42^\circ$ and $48^\circ$. This is because of the fact the neutrino probability is smaller in IH for neutrinos as compared to NH. At both the above mentioned true values, the best octant sensitivity comes for 3+3 years combination but the octant $\chi^2$ does not rise above 3. 

\begin{figure}[htb]
\begin{center}
\includegraphics[width=0.42\textwidth]{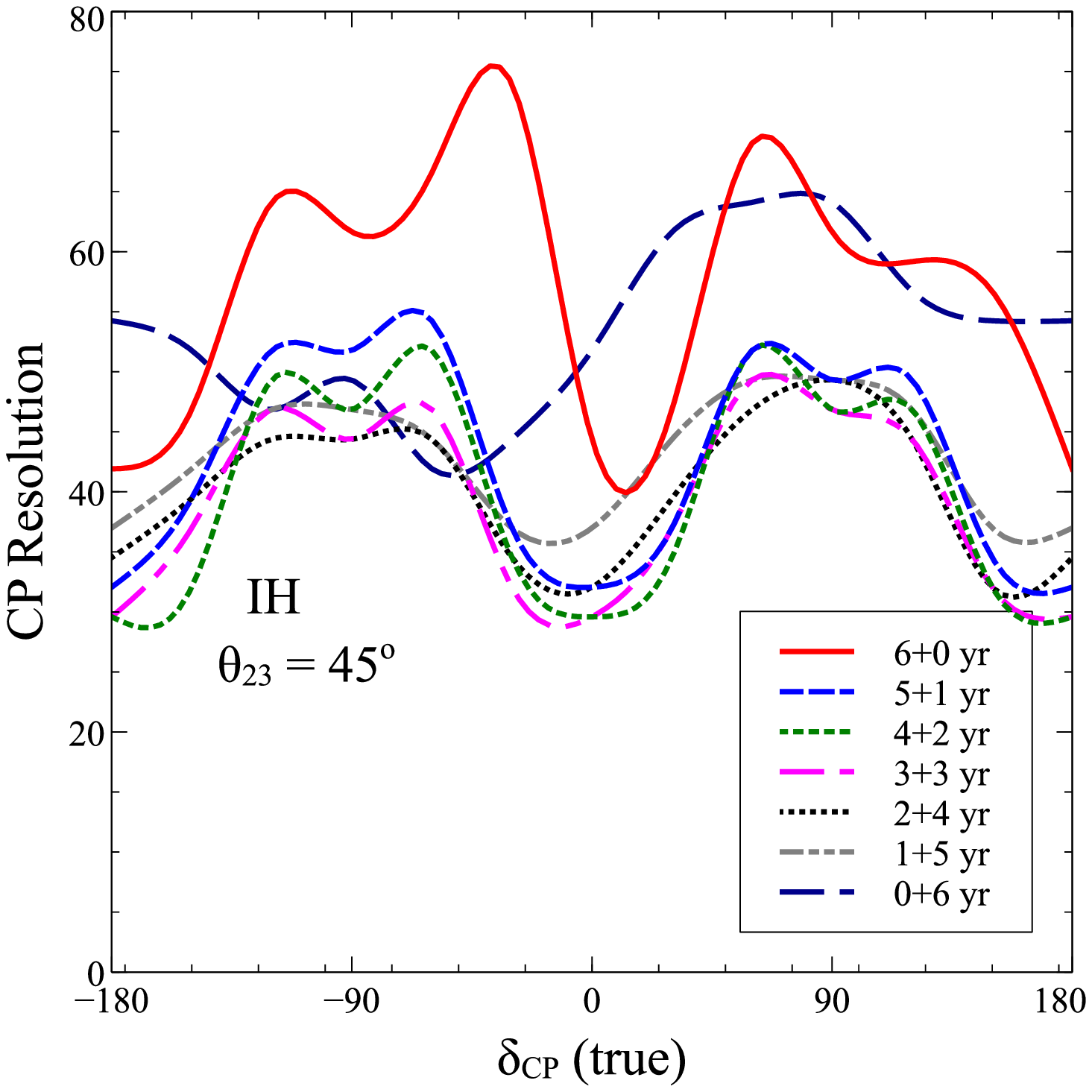}
\end{center}
\caption{CP resolution sensitivity in IH as a function of $\dcp$(true) for $\theta_{23}=45^\circ$. This is presented for four different neutrino-antineutrino ratios.
}
\label{cpp_IH}
\end{figure}
We have presented the results for CPV discovery potential of P2O for different neutrino and antineutrino run-times for IH in fig.~\ref{cpv_IH} with a total run-time of 6 years and with $\theta_{23}=45^{\circ}$. In the left panel we have given the CPV discovery sensitivity as a function of $\dcp({\rm true})$. As for the case of mass hierarchy, here also we have considered 2+4 years, 1+5 years and 0+6 years of run-times along with 6+0 years, 5+1 years, 4+2 years and 3+3 years run-times. 
In the right panel of Fig. \ref{cpv_IH} we have given the required run-time to obtain a certain CPV discovery sensitivity as a function of background for different values of systematics. This we have done taking equal ratio of neutrinos and antineutrinos. The true value of $\dcp$ is $-90^\circ$. Similar to the NH case, there is no effect of systematics and background to achieve a sensitivity of $1 \sigma$ as it requires around 1 year for all the three curves. But unlike NH, the effect of systematics is prominent in IH to achieve a sensitivity of $2 \sigma$. For the most conservative case (i.e., 10\% systematic uncertainty), time required to gain $2 \sigma$ sensitivity rises from 4 years to 6.5 years when background increases from 5\% to 20\%. 
It is also possible to achieve a $3 \sigma$ sensitivity within 10 year of running if the background is less than 7.5\% and systematics in appearance signal channel is absent.

Finally, in Fig. \ref{cpp_IH} we have given the CP resolution capability in IH for $\theta_{23}=45^\circ$ taking different neutrino and antineutrino combinations for a total run-time of six years. From the figure we see that the best sensitivity is obtained for 3+3 years combination and a value of $\dcp ({\rm true})=-90^\circ$, $0^\circ$ and $+90^\circ$ can be measured within 30 degerees of uncertainty.

\section{Comparison of P2O with Other Future Long-baseline Experiments }
\label{comparison}

In this section we compare the sensitivity of the P2O experiment with the other future long-baseline experiments which are DUNE, T2HK and T2HKK.
\begin{figure*}[htb]
 \begin{center}
\includegraphics[width=0.32\textwidth]{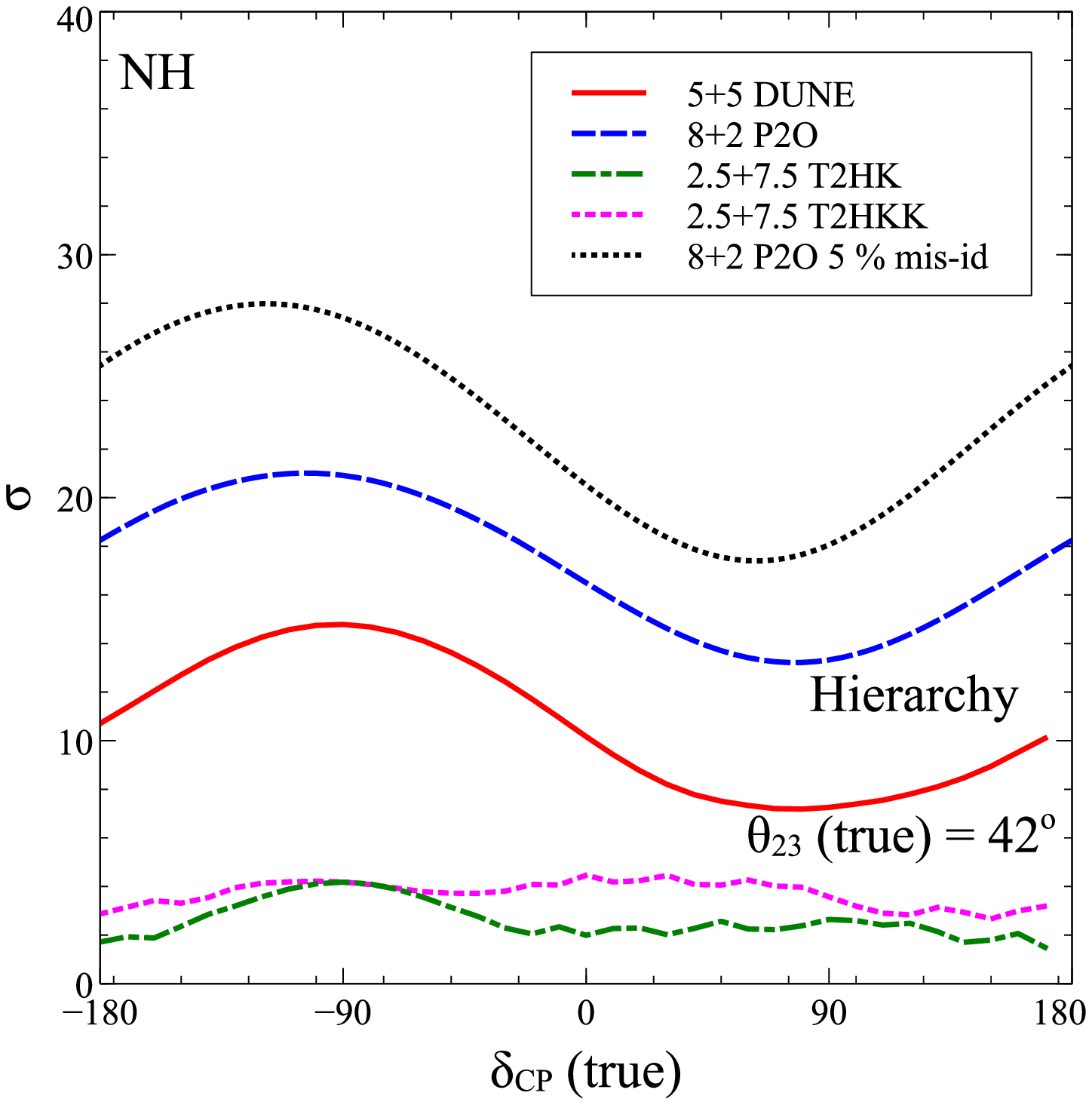} 
\includegraphics[width=0.32\textwidth]{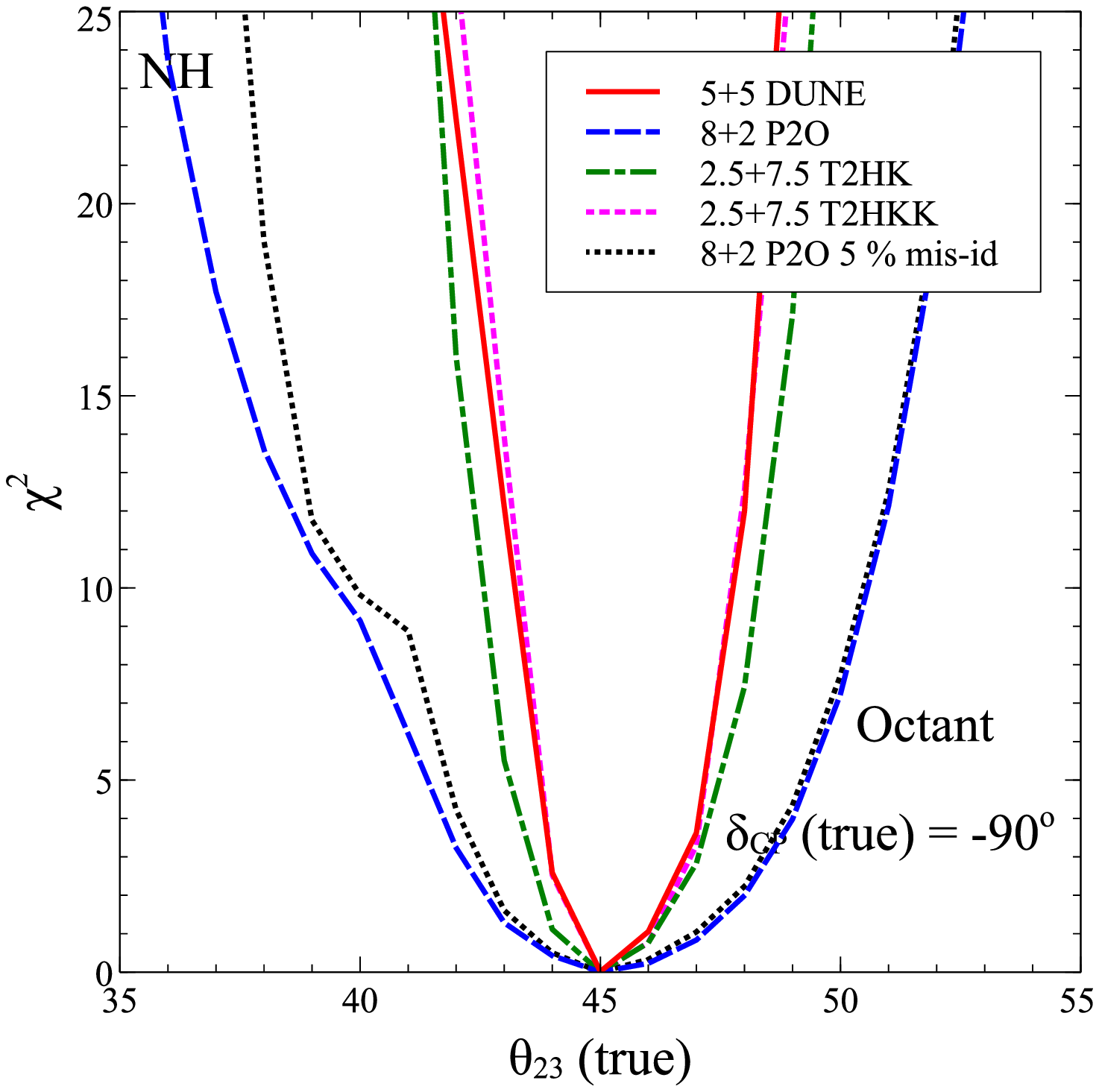}
\includegraphics[width=0.32\textwidth]{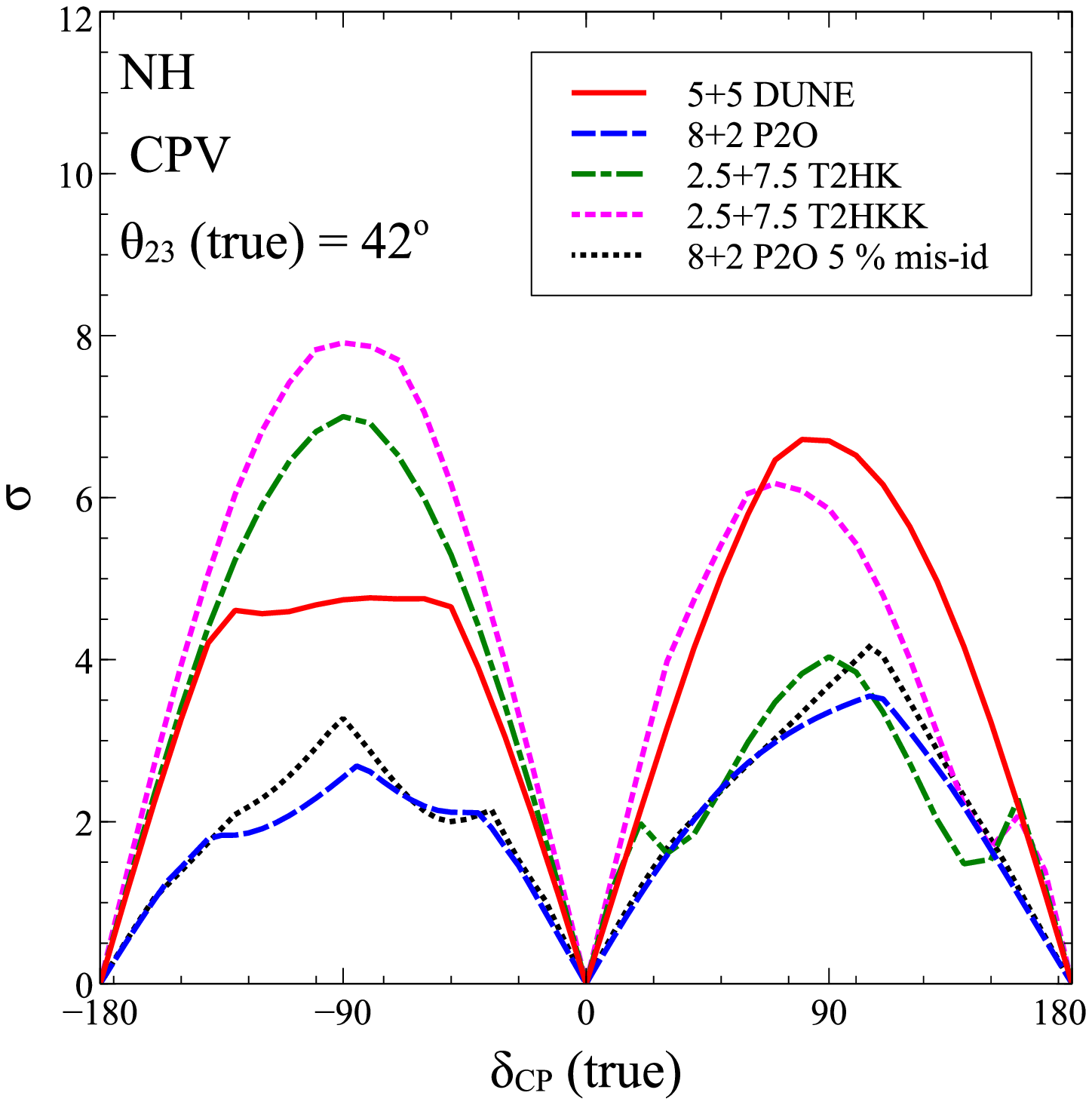} \\
\caption{Comparison of the expected sensitivity of P2O, DUNE, T2HK and T2HKK to discover neutrino mass hierarchy (left panel), octant of $\theta_{23}$ (middle panel) and CPV (right panel). For hierarchy and CPV sensitivity the true value of $\theta_{23}$ is $42^\circ$ and for octant sensitivity the true value of $\dcp$ is $-90^\circ$.}
\label{compare}
 \end{center}
\end{figure*}

In Fig. \ref{compare} we compare the hierarchy, octant and CP violation sensitivity of the above mentioned experiments in the left, middle and right panel, respectively, for NH. We have taken a total run time of 10 years for all the four experiments. As described in the letter of intents, for DUNE (T2HK/T2HKK) we have taken the 1:1 (1:3) ratio for neutrino and antineutrino run. For P2O we have taken the most optimized 8+2 years configuration with two choices of background i.e., 20\% (blue curve) and 5\% (black curve). The choices of true values are mentioned in the figures. For hierarchy sensitivity we see that P2O wins over all the other experiments even with 20\% background. But for octant, P2O has the worst sensitivity among the given experiments. However we checked that with 5\% background and 1\% systematic error, the sensitivity of P2O is comparable to T2HK, for $\theta_{23}$ values closer to maximal. For CPV, P2O and T2HK have comparable sensitivity for $\dcp=+90^\circ$ for 20\% background in P2O, while P2O is better than T2HK for 5\% background. On the other hand, for $\dcp=-90^\circ$, P2O has the worst sensitivity for both the choices of background.

\section{Conclusion}
\label{conclusions}

There is a proposal (P2O) to upgrade the Protvino accelerator complex with an increasing beam power of 450 kW and use it to produce and send a neutrino beam to the ORCA detector in the Mediterranean sea at a distance of 2588 km. The beam will be peaked at about 5 GeV to produce at oscillation maximum at ORCA. The preliminary sensitivity reach of this experiment to neutrino mass hierarchy and CP violation has been presented by the P2O collaboration for 3 years of running of the experiment in the neutrino channel. In this paper, we have performed a detailed sensitivity study of P2O for all three neutrino physics parameters measurable at long-baseline experiments - neutrino mass hierarchy, CP violation and octant of $\theta_{23}$. The optimisation is done with respect to (i) neutrino vs antineutrino run-time, (ii) detector systematic uncertainties, (iii) total running time of the experiment and (iv) backgrounds coming from mis-id in the appearance channel. 

We started by matching our simulated number of events as well a $\chi^2$ with the results presented by the P2O collaboration \cite{p2o_talk_zaborov} for 3 years run of the experiment in the neutrino channel and for NH true. We next calculated the corresponding events for the antineutrinos using fluxes obtained from the P2O collaboration, using the same simulation parameters as for the neutrino channel. We assumed a total run time of 6 years for the experiment and varied the neutrino vs antineutrino run time ratio for mass hierarchy, CP violation and octant of $\theta_{23}$. We showed that for mass hierarchy measurement, the sensitivity of P2O is always very high for both hierarchies. While the sensitivity does change with neutrino-antineutrino fraction, systematics and backgrounds, especially for NH true. The significance with which mass hierarchy can be measured remains high for both NH and IH true.

The situation with CP violation and octant of $\theta_{23}$ determination is more complicated for P2O. We found that the significance of CP violation measurement for the baseline design of the experiment is not as high as that for DUNE and T2HK. For the current best-fit of $\delta_{CP}=-90^\circ$ and $\theta_{23}$(true)$=42^\circ$, it can be measured at $2\sigma$ for 5+1 years of running of the experiment with 20\% mis-id background. We showed that this could be increased to $3\sigma$ if the background is reduced to to 5\% and  run time increased to 8+2 years. For octant the reach of P2O baseline design is expected to be less than $2\sigma$ at $\theta_{23}$(true)$=42^\circ$ and $\delta_{CP}=-90^\circ$ with a background of 20\%. We showed that if the mis-id is controlled within 5\%, then octant can be measured with $\chi^2=5$ for 8+2 years run of the experiment. Finally, we made a comparative study of P2O along with DUNE, T2HK and T2HKK and showed that P2O is expected to give the best sensitivity to neutrino mass hierarchy owing to its long baseline which is close to being bi-magic. In terms of CP violation and octant of $\theta_{23}$ discovery, even though P2O baseline configuration appears to have a weaker sensitivity compared to DUNE, T2HK and T2HKK, we showed what is needed in order for P2O to be competitive with other long-baseline proposals. 

In conclusion, P2O has the best sensitivity to neutrino mass hierarchy. For octant of $\theta_{23}$ and CP violation discovery it can be competitive if the experimental collaboration is able to control mis-id and systematics, and run the experiment for 8+2 years in neutrino and antineutrino mode.

 Note that there is a propsal for a super-ORCA detector with approximately 10 times more densely instrumented version of ORCA with lower detection threshold and improved event reconstruction capabilities \cite{p2o_poster}. It was shown that with this upgraded detector, $\dcp = 0^\circ$ and $180^\circ$ can be distinguished with $5 \sigma$ confidence level and  60\% (70\%) of $\dcp$ values can be disfavoured by more than $2 \sigma$ confidence level for true $\dcp =0^\circ$, $180^\circ$ ($\pm 90^\circ$).

\begin{acknowledgements}

The authors would like to thank Prof. Dmitry Zaborov and Prof. Anatoly Sokolov of the P2O collaboration for providing the fluxes and relevant information regarding the P2O experiment.  The authors would like to thank the Department of Atomic Energy (DAE) Neutrino Project of Harish-Chandra Research Institute. This project has received funding from the European Union's Horizon 2020 research and innovation programme InvisiblesPlus RISE under the Marie Sklodowska-Curie grant agreement No 690575. This project has received funding from the European Union's Horizon 2020 research and innovation programme Elusives ITN under the Marie Sklodowska- Curie grant agreement No 674896.

\end{acknowledgements}

\bibliographystyle{h-elsevier}
\bibliography{neutosc}   

\begin{thebibliography}{10}

\bibitem{Acciarri:2015uup}
DUNE, R. Acciarri et~al.,
\newblock (2015), 1512.06148.

\bibitem{Abe:2016ero}
Hyper-Kamiokande, K. Abe et~al.,
\newblock PTEP 2018 (2018) 063C01, 1611.06118.

\bibitem{Baussan:2013zcy}
ESSnuSB, E. Baussan et~al.,
\newblock Nucl. Phys. B885 (2014) 127, 1309.7022.

\bibitem{Agarwalla:2017nld}
S.K. Agarwalla, M. Ghosh and S.K. Raut,
\newblock JHEP 05 (2017) 115, 1704.06116.

\bibitem{Ghosh:2017ged}
M. Ghosh and O. Yasuda,
\newblock Phys. Rev. D96 (2017) 013001, 1702.06482.

\bibitem{Fukasawa:2016yue}
S. Fukasawa, M. Ghosh and O. Yasuda,
\newblock Nucl. Phys. B918 (2017) 337, 1607.03758.

\bibitem{Nath:2015kjg}
N. Nath, M. Ghosh and S. Goswami,
\newblock Nucl. Phys. B913 (2016) 381, 1511.07496.

\bibitem{Ghosh:2014rna}
M. Ghosh, S. Goswami and S.K. Raut,
\newblock Eur. Phys. J. C76 (2016) 114, 1412.1744.

\bibitem{Chakraborty:2017ccm}
K. Chakraborty, K.N. Deepthi and S. Goswami,
\newblock Nucl. Phys. B937 (2018) 303, 1711.11107.

\bibitem{Raut:2017dbh}
S.K. Raut,
\newblock Phys. Rev. D96 (2017) 075029, 1703.07136.

\bibitem{Barger:2014dfa}
V. Barger et~al.,
\newblock Int. J. Mod. Phys. A31 (2016) 1650020, 1405.1054.

\bibitem{Barger:2013rha}
V. Barger et~al.,
\newblock Phys. Rev. D89 (2014) 011302, 1307.2519.

\bibitem{Agarwalla:2014tpa}
S.K. Agarwalla, S. Choubey and S. Prakash,
\newblock JHEP 12 (2014) 020, 1406.2219.

\bibitem{Chatterjee:2017irl}
S.S. Chatterjee, P. Pasquini and J.W.F. Valle,
\newblock Phys. Rev. D96 (2017) 011303, 1703.03435.

\bibitem{Chatterjee:2017xkb}
S.S. Chatterjee, P. Pasquini and J.W.F. Valle,
\newblock Phys. Lett. B771 (2017) 524, 1702.03160.

\bibitem{Blennow:2014sja}
M. Blennow, P. Coloma and E. Fernandez-Martinez,
\newblock JHEP 03 (2015) 005, 1407.3274.

\bibitem{Ballett:2016daj}
P. Ballett et~al.,
\newblock Phys. Rev. D96 (2017) 033003, 1612.07275.

\bibitem{Evslin:2015pya}
J. Evslin, S.F. Ge and K. Hagiwara,
\newblock JHEP 02 (2016) 137, 1506.05023.

\bibitem{p2o_talk_zaborov}
Scientific potential of a neutrino beam from Protvino to ORCA (P2O), D.
  Zaborov,
\newblock 2017,
\newblock Talk given at Neutrino GDR meeting Paris, France, Nov 20-21, 2017.

\bibitem{Brunner:2013lua}
J. Brunner,
\newblock (2013), 1304.6230.

\bibitem{Zaborov:2018whl}
KM3NeT, D. Zaborov,
\newblock {18th Lomonosov Conference on Elementary Particle Physics Moscow,
  Russia, August 24-30, 2017}, 2018, 1803.08017.

\bibitem{Adrian-Martinez:2016fdl}
KM3Net, S. Adrian-Martinez et~al.,
\newblock J. Phys. G43 (2016) 084001, 1601.07459.

\bibitem{Akindinov:2019flp}
A.V. Akindinov et~al.,
\newblock (2019), 1902.06083.

\bibitem{p2o_beam}
A. Zaitsev,
\newblock 2018,
\newblock Talk given at Very Large Volume Neutrino Telescopes, 2 - 4 October
  2018, Dubna Russia.

\bibitem{Dighe:2010js}
A. Dighe, S. Goswami and S. Ray,
\newblock Phys. Rev. Lett. 105 (2010) 261802, 1009.1093.

\bibitem{Raut:2009jj}
S.K. Raut, R.S. Singh and S.U. Sankar,
\newblock Phys. Lett. B696 (2011) 227, 0908.3741.

\bibitem{globes}
P. Huber, M. Lindner and W. Winter,
\newblock Comput. Phys. Commun. 167 (2005) 195, hep-ph/0407333.

\bibitem{Esteban:2018azc}
I. Esteban et~al.,
\newblock (2018), 1811.05487.

\bibitem{deSalas:2017kay}
P.F. de~Salas et~al.,
\newblock Phys. Lett. B782 (2018) 633, 1708.01186.

\bibitem{Capozzi:2018ubv}
F. Capozzi et~al.,
\newblock Prog. Part. Nucl. Phys. 102 (2018) 48, 1804.09678.

\bibitem{Prakash:2012az}
S. Prakash, S.K. Raut and S.U. Sankar,
\newblock Phys. Rev. D86 (2012) 033012, 1201.6485.

\bibitem{Ghosh:2015ena}
M. Ghosh et~al.,
\newblock Phys. Rev. D93 (2016) 013013, 1504.06283.

\bibitem{Ghosh:2015tan}
M. Ghosh,
\newblock Phys. Rev. D93 (2016) 073003, 1512.02226.

\bibitem{Prakash:2013dua}
S. Prakash, U. Rahaman and S.U. Sankar,
\newblock JHEP 07 (2014) 070, 1306.4125.

\bibitem{Agarwalla:2013ju}
S.K. Agarwalla, S. Prakash and S.U. Sankar,
\newblock JHEP 07 (2013) 131, 1301.2574.

\bibitem{p2o_poster}
https://zenodo.org/record/1292936, J. Hofestädt,
\newblock 2018,
\newblock Psoter given at Neutrino 2018, Heidelberg, June 4-9, 2018.

\end{thebibliography}

%
%

\end{document}